\documentclass[prb,citeautoscript,amsmath,twocolumn,amssymb,superscriptaddress]{revtex4-2}
\usepackage{float}
\usepackage[utf8]{inputenc}
\usepackage[T1]{fontenc}
\usepackage{xspace}
\usepackage{physics}
\usepackage{graphicx}
\usepackage[dvipsnames]{xcolor}
\usepackage[separate-uncertainty=true,multi-part-units=single]{siunitx}
\usepackage[version=4]{mhchem}
\usepackage{xr-hyper}
\usepackage{gensymb}
\usepackage{soul}
\usepackage[colorlinks, citecolor={blue}, linkcolor={black}, urlcolor={blue}]{hyperref}
\AtBeginDocument{\usepackage{booktabs}}

\renewcommand{\figurename}{\textbf{\figlbl}} 

\newcommand{\figtitle}[1]{\textbf{#1}\xspace} 

\newcommand{\placeholder}[1]{}

\begin{document}

\title{Electronic ratchet effect in a moir\'e system: signatures of excitonic ferroelectricity}

\author{Zhiren Zheng}
\affiliation{Department of Physics, Massachusetts Institute of Technology, Cambridge, MA, USA}
\author{Xueqiao Wang}
\affiliation{Department of Physics, Massachusetts Institute of Technology, Cambridge, MA, USA}
\author{Ziyan Zhu}
\affiliation{Stanford Institute of Materials and Energy Science, SLAC National Accelerator Laboratory, Menlo Park, CA 94025, USA}
\author{Stephen Carr}
\affiliation{Brown Theoretical Physics Center and Department of Physics, Brown University, Providence, RI, USA}
\author{Trithep Devakul}
\affiliation{Department of Physics, Massachusetts Institute of Technology, Cambridge, MA, USA}
\author{Sergio de la Barrera}
\affiliation{Department of Physics, Massachusetts Institute of Technology, Cambridge, MA, USA}
\author{Nisarga Paul}
\affiliation{Department of Physics, Massachusetts Institute of Technology, Cambridge, MA, USA}
\author{Zumeng Huang}
\affiliation{Department of Physics, Boston College, Chestnut Hill, MA, USA}
\author{Anyuan Gao}
\affiliation{Department of Chemistry and Chemical Biology, Harvard University, Cambridge, MA, USA}
\author{Yang Zhang}
\affiliation{Department of Physics, Massachusetts Institute of Technology, Cambridge, MA, USA}
\author{Damien Bérubé}
\affiliation{Department of Chemistry and Chemical Biology, Harvard University, Cambridge, MA, USA}
\author{Kathryn Natasha Evancho}
\affiliation{Department of Physics and Astronomy, Clemson University, Clemson, South Carolina, USA}
\author{Kenji Watanabe}
\affiliation{Research Center for Functional Materials, National Institute for Materials Science, Tsukuba, Japan}
\author{Takashi Taniguchi}
\affiliation{International Center for Material Nanoarchitectonics, National Institute for Materials Science, Tsukuba, Japan}
\author{Liang Fu}
\affiliation{Department of Physics, Massachusetts Institute of Technology, Cambridge, MA, USA}
\author{Yao Wang}
\affiliation{Department of Physics and Astronomy, Clemson University, Clemson, South Carolina, USA}
\author{Su-Yang Xu}
\affiliation{Department of Chemistry and Chemical Biology, Harvard University, Cambridge, MA, USA}
\author{Efthimios Kaxiras}
\affiliation{Department of Physics, Harvard University, Cambridge, MA, USA}
\author{Pablo Jarillo-Herrero}
\email{pjarillo@mit.edu}
\affiliation{Department of Physics, Massachusetts Institute of Technology, Cambridge, MA, USA}
\author{Qiong Ma}
\email{qiong.ma@bc.edu}
\affiliation{Department of Physics, Boston College, Chestnut Hill, MA, USA}
\affiliation{CIFAR Azrieli Global Scholars program, CIFAR, Toronto, Canada}

\begin{abstract}
Electronic ferroelectricity represents a new paradigm where spontaneous symmetry breaking driven by electronic correlations, in contrast to traditional lattice-driven ferroelectricity, leads to the formation of electric dipoles. Despite the potential application advantages arising from its electronic nature, switchable electronic ferroelectricity remains exceedingly rare. Here, we report the discovery of an electronic ratchet effect that manifests itself as switchable electronic ferroelectricity in a layer-contrasting graphene-boron nitride moir\'e heterostructure. Our engineered layer-asymmetric moir\'e potential landscapes result in layer-polarized localized and itinerant electronic subsystems. At particular fillings of the localized subsystem, we find a ratcheting injection of itinerant carriers in a non-volatile manner, leading to a highly unusual ferroelectric response. Strikingly, the remnant polarization can be stabilized at multiple (quasi-continuous) states with behavior markedly distinct from known ferroelectrics. Our experimental observations, simulations, and theoretical analysis suggest that dipolar excitons are the driving force and elementary ferroelectric units in our system. This signifies a new type of electronic ferroelectricity where the formation of dipolar excitons with aligned moments generates a macroscopic polarization and leads to an electronically-driven ferroelectric response, which we term excitonic ferroelectricity. Such new ferroelectrics, driven by quantum objects like dipolar excitons, could pave the way to innovative quantum analog memory and synaptic devices.

\end{abstract}

\maketitle

\subsection*{Introduction}

Electronic ferroelectricity stands as a counterpart to traditional lattice-driven ferroelectricity where the formation of electric dipoles arises from electronic degrees of freedom instead of ionic displacement. The realization of such a phase of quantum matter could enable the exploration of a highly sought-after parameter space regime with intertwined correlations, spatial symmetry breaking, and Berry phase ~\cite{ishihara2010electronic, yamauchi2014electronic}. Additionally, electronic ferroelectricity is expected to exhibit novel characteristics, such as ultrafast switching, large polarization, exceptional durability, and multiple polarization states~\cite{tokura2017emergent}. However, candidates for electronic ferroelectricity are extremely rare, confined to a handful of oxides and certain organic salts~\cite{yamauchi2014electronic}. Moreover, due to the weakly insulating nature of electronic ferroelectrics, traditional detection and control schemes for ferroelectric insulators are not well-suited for their study. Hence, experimental demonstration of switchable electronic ferroelectricity is challenging and remains elusive.

As one of the very few concrete proposals for electronic ferroelectricity~\cite{ishihara2010electronic, yamauchi2014electronic}, exciton-driven ferroelectricity is an intriguing yet underexplored concept. An exciton is a Coulomb-bound state between a positive charge (hole) and a negative charge (electron). A dipolar exciton, where the electron and hole are spatially separated, possesses an electric dipole moment. Consequently, a collection of dipolar excitons with aligned moments results in a macroscopic electric polarization. Despite the intuitive idea, theoretical investigation for exciton-driven ferroelectricity is limited. One example is mixed-valence compounds, proposed by Portengen et al.~\cite{portengen1996theory} in 1996 (called the POS model in the following). It was proposed that spontaneous electric polarization can emerge from coherence coupling between itinerant \textit{d}-electrons and localized \textit{f}-holes through an on-site Coulomb repulsion mechanism~\cite{batyev1980excitonic, portengen1996theory, batista2002electronic}. Another example is bilayer graphene, proposed by Nandkishore et al.~\cite{nandkishore2010dynamical} in 2010. It was predicted that excitonic instability due to electron interaction in undoped bilayer graphene can lead to a ferroelectric state. When quantum objects like excitons become individual electric dipoles, fundamentally new phenomena and functionalities are expected that could lead to novel quantum devices for neuromorphic computing~\cite{christensen20222022}. Despite its great fundamental and technological prospects, exciton-driven ferroelectricity remains experimentally unexplored.

Moir\'e quantum materials constitute a highly tunable and configurable material platform, providing an opportunity to realize exciton-driven ferroelectricity. Recent studies have shown that moir\'e engineered flat band systems can exhibit strong electron interactions and Coulomb-driven phases that give rise to spontaneous electronic orderings~\cite{cao2018correlated, cao2018unconventional, andrei2020graphene, andrei2021marvels, mak2022semiconductor}. The resulting ordering could potentially develop a finite electric polarization, leading to a ferroelectric ground state. In addition, the prevailing existence of interlayer excitons and strong exciton interactions~\cite{rickhaus2021correlated, gu2022dipolar, chen2022excitonic, zhang2022correlated} in various layered moir\'e systems  further enriches the phase diagram and offers possibilities for realizing exciton-driven ferroelectricity. Due to the rarity of electronic ferroelectricity, let alone exciton-driven ferroelectricity, there is currently no established standard protocol for its demonstration.

In this work, we investigate Bernal bilayer graphene (BLG) sandwiched between top and bottom boron nitride (BN) layers at vastly different rotational angle alignments (one BLG-BN twist angle is small and the other BLG-BN twist angle is large). The large angle mismatch between the top and bottom BNs explicitly breaks the lattice inversion symmetry of the system. We note however that, despite the fixed lattice asymmetry, we observe a novel electronic ratchet effect that induces spontaneous remnant electric polarization with equal and opposite configurations, illustrating the importance of the electronic degrees of freedom. We further demonstrate that the remnant polarization can be stabilized in a quasi-continuous fashion. This unique ferroelectric response can be described by our phenomenological model based on discrete ferroelectric units. Our experimental observations and theoretical investigation is consistent with a scenario where dipolar excitons serve as the driving force and the elementary electric dipole units of the ferroelectric response, which motivates us to use the name excitonic ferroelectricity. Despite some similar underlying ingredients, our experimental findings reside in a different regime from previously proposed exciton-driven ferroelectricity~\cite{portengen1996theory}.
Below we show first our experimental data, which we have obtained consistently in multiple devices (we present data measured in Device D1 in full detail, see Extended data Fig.~\ref{hyssummary} for other devices). After that we show our theoretical modeling.

\subsection*{Basic transport characterization}

Due to their lattice mismatch, graphene and BN form a long-range moir\'e superlattice with a moir\'e wavelength of about 14 nm even when they are perfectly aligned. As the rotational angle is tuned away from zero, the moir\'e wavelength decreases gradually. Our device design consists of BLG closely aligned with the top BN, and purposely misaligned with the bottom BN by 15$^{\circ}$. This way each graphene sheet in the BLG experiences a distinct moir\'e potential landscape ~\cite{yankowitz2019van}, as shown in Figs.~\ref{Fig1}\textbf{a-b} (see Methods and Extended data Fig.~\ref{SHG} for careful angle alignment and determination). The real-space wavefunction distribution, obtained from single-particle band structure calculations, displays a long wavelength modulation at the top moir\'e interface and uniform wavefunction distribution at the bottom moir\'e interface (see Methods and Extended data Fig.~\ref{calculation})~\cite{zhu2022electric}. We note that, although the concept of layer-specific electronic structure in graphene-BN moir\'e structures has not been investigated in detail, this idea can be generalized to other layered moir\'e systems and it might become a key aspect in the understanding of a variety of recently observed phenomena~\cite{rozen2021entropic, saito2021isospin, park2021tunable, hao2021electric, zhao2022gate, ramires2021emulating, song2022magic, kumar2022gate, zeng2022layer}. This unique electronic structure has significant effects on transport characteristics.

Figure~\ref{Fig1}\textbf{c} shows a measurement of the four-probe resistance as a function of bottom and top gate voltages ($V_\mathrm{BG}$, $V_\mathrm{TG}$), respectively. In regular bilayer graphene, the resistance maximum traces the charge neutrality point, where the total charge density $n \propto\frac{V_\mathrm{BG}}{d_\mathrm{b}}+\frac{V_\mathrm{TG}}{d_\mathrm{t}}=0$ ($d_\mathrm{t(b)}$ is the thicknesses of the top (bottom) BN). The resistance peak is expected to follow a straight diagonal line whose slope is determined by the ratio between $d_\mathrm{t}$ and $d_\mathrm{b}$~\cite{taychatanapat2010electronic}. In sharp contrast, Device D1 exhibits distinct regions where the maximum resistance follows lines with different slopes. In regions \textbf{ii} and \textbf{iv}, the resistance maximum follows a slope determined by the ratio of the BN thicknesses (black dashed lines in Fig.~\ref{Fig1}\textbf{c}), as it might be expected. However, we note that these regions exhibit very unusual charge dynamics, which will be discussed below. We denote regions \textbf{ii} and \textbf{iv} as the ratcheting regimes. In regions \textbf{i}, \textbf{iii}, and \textbf{v}, on the other hand, the resistance maximum follows slopes that deviate from the $d_\mathrm{t}/d_\mathrm{b}$ ratio  expectation~\cite{zheng2020unconventional, klein2023electrical,  niu2022giant}. In these regions, $V_\mathrm{TG}$ is ``anomalously" ineffective at charging the system, as if its electric field is screened. Hence we denote these regions as the Layer-Specific Anomalous Screening (LSAS) regimes. Most dramatically, in region \textbf{iii}, the effect of $V_\mathrm{TG}$ is completely canceled so that the resistance versus $V_\mathrm{BG}$ curves are independent of $V_\mathrm{TG}$ (inset of Fig.~\ref{Fig1}\textbf{c}). Overall, the alternating behavior between the LSAS and ratcheting regimes forms a cascade pattern.

\subsection*{Coexistence of itinerant and localized systems}

\begin{figure*}
    \includegraphics[width=4in]{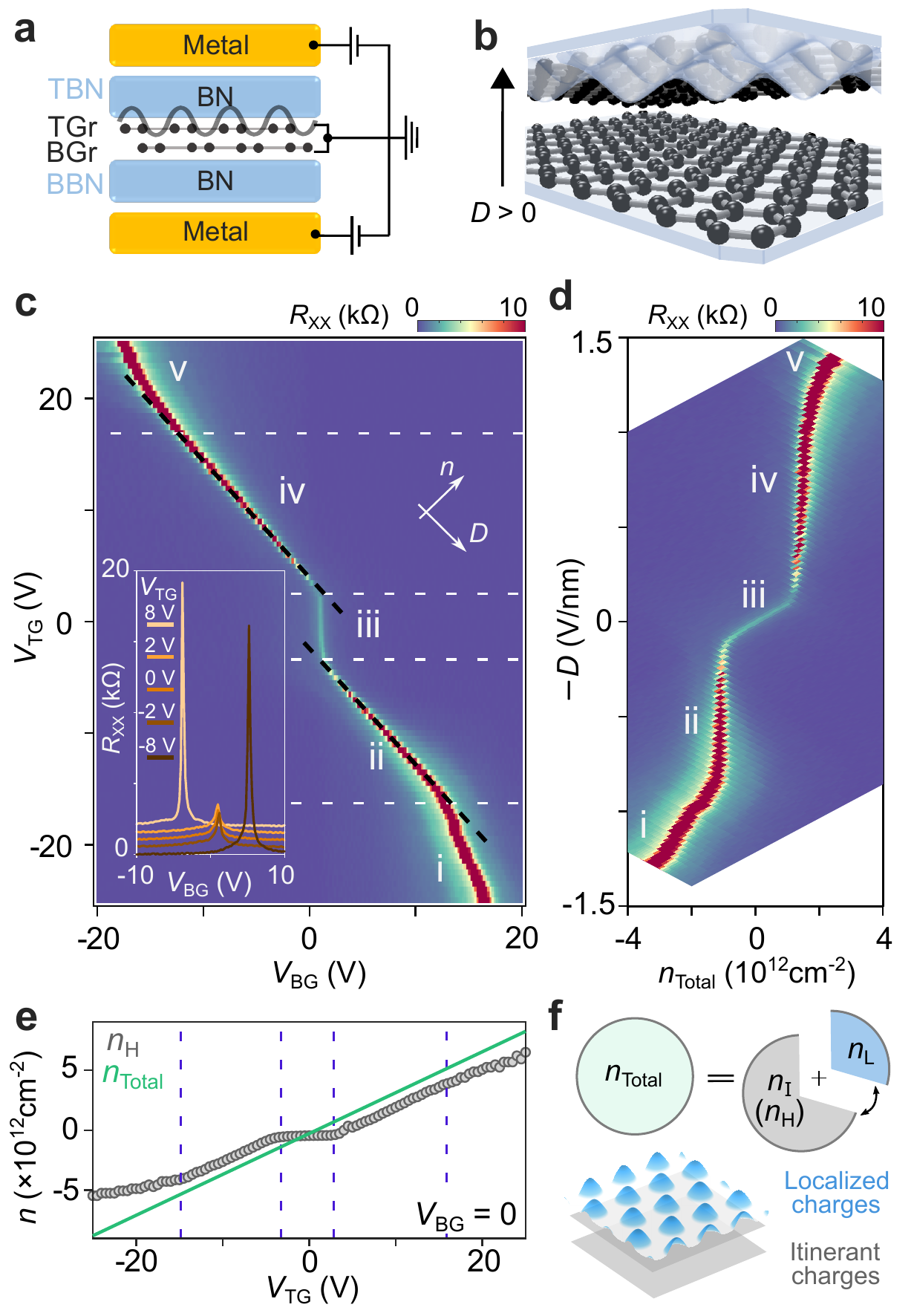}
    \caption{\figtitle{Layer-specific moir\'e potential and the layer-specific anomalous screening.}
    (\textbf{a}) Bernal stacked bilayer graphene encapsulated in top and bottom BN in a dual-gated device structure. (\textbf{b}) The asymmetric moir\'e potential gives rise to layer-specific potential modulation on the top and bottom layers of graphene. The schematics show a representative scenario where the bilayer graphene is closely aligned to the top BN and misaligned with the bottom BN at a large angle. (\textbf{c}) Four-probe resistance as a function of $V_\mathrm{BG}$ and $V_\mathrm{TG}$. Inset: line cuts taken at fixed $V_\mathrm{TG}$. $V_\mathrm{BG}$ can always significantly modulate the conductance, while $V_\mathrm{TG}$ has minimal effect on the device conductance for voltage ranging from $-3$ to $3$ V. (\textbf{d}) Converted resistance map as a function of density $n_\mathrm{Total}$ versus the displacement field $D$. (\textbf{e}) The grey curve shows the Hall density $n_\mathrm{H}$ as a function of $V_\mathrm{TG}$ at $V_\mathrm{BG} = 0$. The green curve shows the gate-induced total density $n_\mathrm{Total}$ calculated based on the filling efficiency in the ratcheting regime $n_\mathrm{total} \sim 0.34 \times (V_\mathrm{BG}+V_\mathrm{TG})$ ($10^{12}$ cm$^{-2})$. (\textbf{f}) Schematic diagram that illustrates the gate-induced total density as a sum of localized and itinerant charge densities.
    }
    \label{Fig1}
\end{figure*}

Our symmetry analysis and Hall measurements below reveal that the LSAS regime arises from layer-polarized charge localization due to the layer-contrasting moir\'e potential. By converting the $V_\mathrm{BG}$-$V_\mathrm{TG}$ resistance map to a density($n$)-displacement field($D$) resistance map (where $D=\frac{\epsilon_\mathrm{BN}}{2}(V_\mathrm{BG}/d_\mathrm{b}-V_\mathrm{TG}/d_\mathrm{t})$) (Fig.~\ref{Fig1}\textbf{d}), we illustrate the unique $n \times D$ symmetry of the system. Evidently, the cascade pattern strongly breaks the symmetry between electrons and holes ($n$ to $-n$) and the symmetry between positive and negative displacement fields ($D$ to $-D$). However, the symmetry is approximately preserved under a combined ($n$ to $-n$) and ($D$ to $-D$) operation. This unique symmetry of the complex cascade pattern, which has been consistently obtained across multiple samples (Extended data Fig.~\ref{hyssummary}), is very unlikely to be due to extrinsic effects (see SI.C.1) and strongly supports the intrinsic nature of the screening behavior. In fact, the symmetry aligns well with the layer-contrasting moir\'e structure, i.e., at $D>0$ $(D<0)$, holes (electrons) are polarized towards the top layer and hence experience the long-range moir\'e potential.

To gain further insight into the type of charge carriers induced in each regime, we measure the Hall density, which characterizes the density of mobile or itinerant carriers. Our measurements reveal the coexistence of an itinerant subsystem and a layer-specific localized subsystem. Figure~\ref{Fig1}\textbf{e} shows the measured Hall density ($n_\mathrm{H}$) as a function of $V_\mathrm{TG}$ for a fixed $V_\mathrm{BG} = 0$ V. Consistent with Fig.~\ref{Fig1}\textbf{c}, $V_\mathrm{TG}$ is ``anomalously" ineffective at charging the system in regions \textbf{i}, \textbf{iii}, and \textbf{v}. The green curve in Fig.~\ref{Fig1}\textbf{e} shows the calculated total charge density ($n_\mathrm{Total}$) induced by the electrostatic gating. The deviation of $n_\mathrm{H}$ from $n_\mathrm{Total}$ reveals the presence of ``hidden'' (or effectively localized) charges that are not contributing to $n_\mathrm{H}$. Their origin arises from the presence of the top moir\'e potential, which can lead to charge localization within an array of moir\'e sites at the top interface. This picture is supported by our single-particle calculations mentioned earlier. Hence, $n_\mathrm{Total}$ can be considered as the sum of the itinerant charge density $n_\mathrm{I}$ (characterized by $n_\mathrm{H}$), which is regular BLG-like, and the intrinsic localized charge density $n_\mathrm{L}$ at the top moir\'e interface (Fig.~\ref{Fig1}\textbf{f}). The itinerant and localized subsystems, in parallel, contribute to electron transport, where the conductance is dominated by the highly conductive itinerant subsystem. The LSAS is a natural consequence of $V_\mathrm{TG}$ charging the top interface, where the localized electronic system resides.

\subsection*{Characterization of the localized system}

\begin{figure*}
    \includegraphics[width=6.5in]{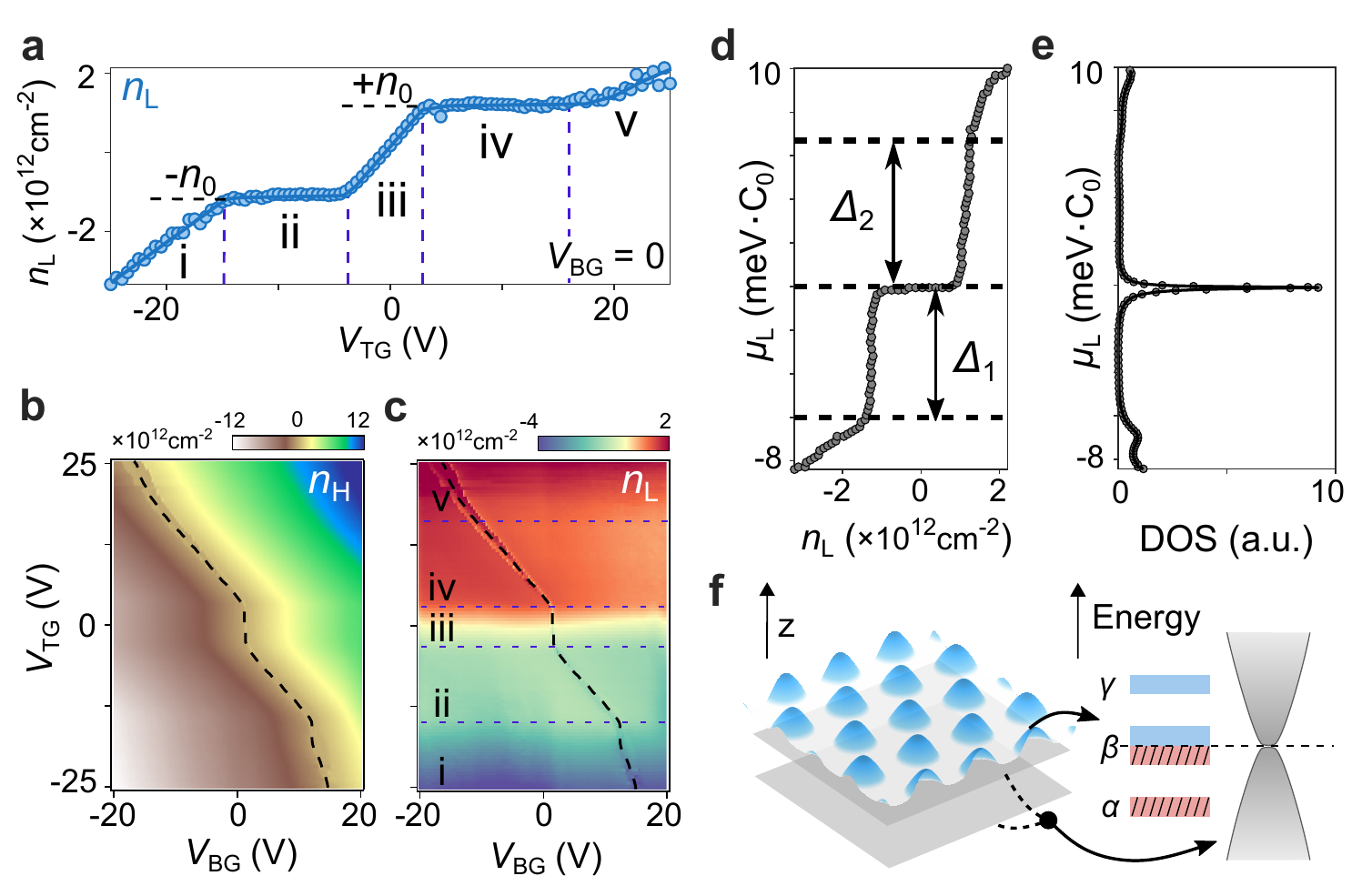}
    \caption{\figtitle{Extraction of the carrier density and chemical potential of the localized subsystem.}
    (\textbf{a}) Localized charge density $n_\mathrm{L}$ calculated as a function of $V_\mathrm{TG}$ at $V_\mathrm{BG} = 0$. (\textbf{b}) Experimentally measured $n_\mathrm{H}$ map as a function of $V_\mathrm{BG}$ and $V_\mathrm{TG}$ with full gate range. (\textbf{c}) $n_\mathrm{L}$ calculated for the full gate range. (\textbf{d}) Chemical potential of the localized subsystem $\mu_\mathrm{L}$ extracted as a function of its density $n_\mathrm{L}$. $\mu_\mathrm{L}$ and $n_\mathrm{L}$ are extracted self-consistently through electrostatic relations (see Methods). (\textbf{e}) Density of states (DOS), defined as $\frac{dn_\mathrm{L}}{d\mu_\mathrm{L}}$, calculated as a function of the localized subsystem's chemical potential $\mu_\mathrm{L}$. (\textbf{f}) The localized charges are spatially confined at the top moir\'e interface, while the itinerant charges are distributed between the two graphene sheets. The separation of the two subsystems in real space can be mapped to two sets of coexisting electronic bands in momentum space, of which one corresponds to a localized electron band and the other to a BLG-like dispersive band.
    }
    \label{Fig2}
\end{figure*}

Following the above analysis, we can extract $n_\mathrm{L}$ based on $n_\mathrm{L} = n_\mathrm{Total} - n_\mathrm{H}$. Figure~\ref{Fig2}\textbf{a} shows $n_\mathrm{L}$ calculated from the $n_\mathrm{Total}$ and $n_\mathrm{H}$ curves in Fig.~\ref{Fig1}\textbf{e}. $n_\mathrm{L}$ increases in regions \textbf{i}, \textbf{iii}, and \textbf{v}, while it remains constant in regions \textbf{ii} ($n_\mathrm{L} = -n_0$) and \textbf{iv} ($n_\mathrm{L} = +n_0$). Going from $-n_0$ to $+n_0$ (region $\textbf{iii}$), the localized system is capable of storing charges up to $\sim 2.2\times 10^{12}$ cm$^{-2}$. In SI.A, we show examples of additional charging intervals beyond $n_\mathrm{L}=\pm n_0$. By analogy with the consecutive filling of spin-valley flavors in magic-angle graphene, $-n_0$ and $+n_0$ most likely correspond to a density of one hole and one electron per moir\'e site~\cite{zondiner2020cascade, wong2020cascade}, respectively. This designation allows us to determine a rotational angle of $0.94^\circ$ between the top BN and BLG. [We note, however, that this filling allocation does not affect the discussion below]. By applying the same procedure to each $V_\mathrm{BG}$, we obtain a full $V_\mathrm{BG}-$$V_\mathrm{TG}$ map for $n_\mathrm{H}$ (Fig.~\ref{Fig2}\textbf{b}) and, correspondingly, for $n_\mathrm{L}$(Fig.~\ref{Fig2}\textbf{c}). The overall behavior of $n_\mathrm{L}$ is mainly influenced by $V_\mathrm{TG}$, reinforcing our previous argument that the localized subsystem is strongly associated with the top interface.

We can further extract the chemical potential of the localized system ($\mu_\mathrm{L}$) by treating the itinerant system as a chemical potential sensor~\cite{kim2012direct, park2021flavour} (see Methods and Extended data Fig.~\ref{mu}). The result reveals large chemical potential jumps ($\Delta_1$ and $\Delta_2$) between consecutive LSAS regimes, i.e., between regions \textbf{i} and \textbf{iii}, and between regions \textbf{iii} and \textbf{v}, as shown in Fig.~\ref{Fig2}\textbf{d}. By assuming a vacuum gap of 1 $\mathrm{\AA}$ between the localized and itinerant subsystems, we estimate the energy gaps to be $\approx$ 80 meV, which is consistent with mesoscopic \textit{ab initio} simulations (see SI.C.3). The density of states (DOS, defined as $dn_\mathrm{L}/d\mu_\mathrm{L}$) exhibits peaks in the chemical potential ranges that correspond to the LSAS regimes, as shown in Fig.~\ref{Fig2}\textbf{e}. These observations indicate that the electronic structure of the localized subsystem consists of three sets of localized electron states, which are separated by two energy gaps and referred to as the localized bands $\alpha$, $\beta$, and $\gamma$.

So far, our experimental data reveals the coexistence of an itinerant system and a (top interface) localized system, which can be mapped to two distinct electronic structures (see SI.B \& C.2). One has dispersive bands, while the other consists of three localized bands, as shown in Fig.~\ref{Fig2}\textbf{f}. Red and blue colors represent hole-like and electron-like localized bands. The electron-hole symmetry is necessary to account for the $n$-$D$ symmetric cascade pattern.

\subsection*{Electron and hole ratchets - coupling between the itinerant and localized system}

The above analysis of the itinerant system and the (top interface) localized system is a natural consequence of the layer-specific moir\'e potential. The sequential doping of the two systems gives rise to the cascade of LSAS regimes. Below, we reveal a strong coupling between the two subsystems that leads to a novel electronic ratchet effect controlled by $V_\mathrm{TG}$.

\begin{figure*}
    \includegraphics[width=6.5in]{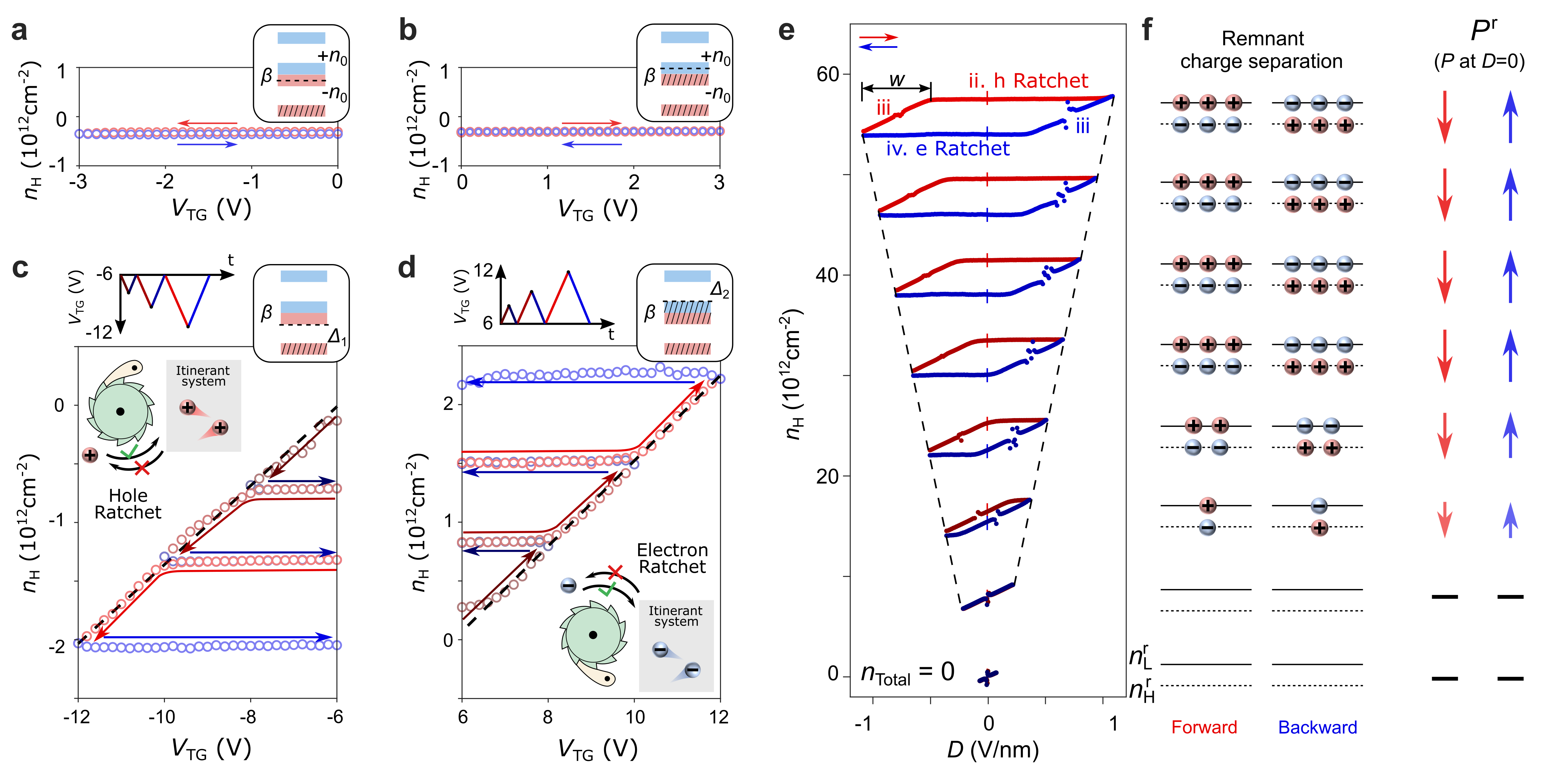}
    \caption{\figtitle{Ratcheting injection of itinerant carriers and scalable ferroelectricity.} Within the localized band $\beta$ ($-n_0<n_\mathrm{L}<+n_0$), $V_\mathrm{TG}$ only induces charges in the localized subsystem, leaving the itinerant charge density unchanged. (\textbf{a} and \textbf{b}) Scanning $V_\mathrm{TG}$ backward and forward within the LSAS regime (region \textbf{iii}) yields a reversible process. However, this process is irreversible when the localized band $\beta$ is depleted or filled. (\textbf{c}) When $n_\mathrm{L}=-n_0$ (region \textbf{ii}), $n_\mathrm{H}$ dependence following the scanning sequence shows a unidirectional hole density increase in the itinerant system, denoted as the hole ratchet. (\textbf{d}) When $n_\mathrm{L}=+n_0$ (region \textbf{iv}), $n_\mathrm{H}$ dependence following the scanning sequence shows a unidirectional electron density increase in the itinerant system, denoted as the electron ratchet. The hysteretic nature of the electron and hole ratchets enables a highly scalable ferroelectric response. (\textbf{e}) $n_\mathrm{H}$ measured along $n_\mathrm{Total}=0$ in the forward and backward direction as a function of $D$ field. Within a small $D$ field range, no hysteresis is observed in region \textbf{iii}. Upon the appearance of regions \textbf{ii} and \textbf{iv} (hole and electron ratchets), a hysteresis loop develops. Due to the ratcheting behavior, the hysteresis loop grows linearly with the displacement field range (black dashed line). (\textbf{f}) At $n_\mathrm{Total}=D=0$, a finite $n_\mathrm{H}$ indicates a spontaneous charge separation in the localized and itinerant subsystems, supported by an intrinsic polarization. The magnitude of $n_\mathrm{H}^\mathrm{r}$ ($n_\mathrm{H}$ at $D=0$) and $n_\mathrm{L}^\mathrm{r}$ ($n_\mathrm{L}$ at $D=0$) at $n_\mathrm{Total}=0$ is proportional to the remnant polarization $P^\mathrm{r}$ ($P$ at $D=0$) of the system.}
    \label{Fig3}
\end{figure*}

We first study the case when $-n_0<n_\mathrm{L}<+n_0$ (region \textbf{iii}). Figures~\ref{Fig3}\textbf{a-b} show that $n_\mathrm{H}$ remains constant in region \textbf{iii} regardless of the  $V_\mathrm{TG}$ scanning direction, indicating that $V_\mathrm{TG}$ does not modify the charge density in the itinerant system within region \textbf{iii}. This process is reversible with no hysteresis.

When $n_\mathrm{L}=-n_0$ (region \textbf{ii}), the charging dynamics by $V_\mathrm{TG}$ changes dramatically and it becomes irreversible. When $V_\mathrm{TG}$ is lowered from -6 to -8 V (the scanning sequence is shown at the top of Fig.~\ref{Fig3}\textbf{c}), $n_\mathrm{H}$
decreases, indicating that holes are added to the itinerant system (Fig.~\ref{Fig3}\textbf{c} brown dots and arrow). However, in the reverse scan (-8 to -6 V), $n_\mathrm{H}$ remains fixed, indicating that holes are not reversibly removed from the itinerant system. As $V_\mathrm{TG}$ is lowered again (-6 to -10 V), the behavior of $n_\mathrm{H}$ has two stages: $n_\mathrm{H}$ first remains constant between -6 to -8 V, and then $n_\mathrm{H}$ decreases between -8 to -10 V. Reversing the scanning direction (-10 to -6 V) yields a constant $n_\mathrm{H}$ again. Overall, at any position along the black dashed line in Fig.~\ref{Fig3}\textbf{c} ($n_\mathrm{L}=-n_0$), lowering $V_\mathrm{TG}$ decreases $n_\mathrm{H}$ while raising $V_\mathrm{TG}$ leaves $n_\mathrm{H}$ unchanged.

Our discussion in Fig.~\ref{Fig2} has demonstrated that the screening of the localized system is responsible for the constant $n_\mathrm{H}$ region. Therefore, the $n_\mathrm{H}$ dependence in Fig.~\ref{Fig3}\textbf{c} indicates that, at $n_\mathrm{L}=-n_0$, lowering $V_\mathrm{TG}$ adds holes to the itinerant system while raising $V_\mathrm{TG}$ removes holes from the localized system, triggering an immediate re-entrance of region \textbf{iii}. Such asymmetric charging and discharging dynamics function as a hole ratchet, by analogy with a mechanical ratchet where the teeth and pawl together provide the ``locking'' mechanism that enforces a unidirectional motion.

When $n_\mathrm{L}=+n_0$ (region \textbf{iv}), the system exhibits a similar ratcheting behavior. Following the scanning sequence at the top of Fig.~\ref{Fig3}\textbf{d}, we observe a continuous unidirectional injection of electrons into the itinerant system, functioning as an electron ratchet. By replacing electrons with holes, we observe the same phenomenology as in Fig.~\ref{Fig3}\textbf{c}.

The observed ratcheting behavior cannot originate from raising the Fermi level (charging) across a trivial single-particle gap. It therefore hints towards an exotic nature of $\Delta_1$ and $\Delta_2$. The electron-hole symmetric behavior in regions \textbf{ii} and \textbf{iv} further indicates that $\Delta_1$ and $\Delta_2$, associated with electron-like and hole-like flat bands, share the same nature.

\subsection*{Continuously scalable ferroelectricity}

The aforementioned electron and hole ratchets together enable a continuously scalable ferroelectric response with switchable dipole moments. To study its behavior, it is natural to measure the system's response as a function of the external displacement field $D$. By varying $V_\mathrm{TG}$ and $V_\mathrm{BG}$ simultaneously, we measure the Hall density $n_\mathrm{H}$ versus $D$ while we keep $n_\mathrm{Total}$ fixed at zero. In the LSAS regime (region \textbf{iii}), both $n_\mathrm{L}$ and $n_\mathrm{H}$ vary with $D$. In the ratcheting regimes (regions \textbf{ii} and \textbf{iv}), both $n_\mathrm{L}$ and $n_\mathrm{H}$ remain constant with $D$.

Within a small $D$ field range (within region \textbf{iii}), scanning forward and backward yields a reversible process, as shown in Fig.~\ref{Fig3}\textbf{e}. Removing the external field ($D=0$) leads to $n_\mathrm{H} = n_\mathrm{L} = 0$. This is expected since $n_\mathrm{Total}=0$. When the $D$ field exceeds region \textbf{iii}, the system's response becomes hysteretic. As a result, we have non-zero remnant itinerant charge density $n_\mathrm{H}^\mathrm{r}$ ($n_\mathrm{H}$ at $D = 0$) and remnant localized charge density $n_\mathrm{L}^\mathrm{r}$ ($n_\mathrm{L}$ at $D = 0$). The low-energy carriers are redistributed between the localized and itinerant systems. The spontaneous charge separation in the z-direction is a response to an internal polarization $P$, a unique trait to metallic ferroelectrics~\cite{fei2018ferroelectric,sharma2019room,de2021direct}. Hence, the remnant charge separation is a consequence of remnant polarization $P^\mathrm{r}$ ($P$ at $D=0$), where $n_\mathrm{H}^\mathrm{r} \propto -P^\mathrm{r}$.

The equal and opposite $n_\mathrm{H}$ values at $D=0$ correspond to opposite real-space charge distribution along the z direction, as illustrated in Fig.~\ref{Fig3}\textbf{f}. Accordingly, the system develops equal and opposite polarization at $D=0$ induced by externally applied $D$ fields. Therefore, the opposite $n_\mathrm{H}^\mathrm{r}$ is the direct evidence of switchable ferroelectricity. The ferroelectric nature can also be confirmed by the gapless point position in the resistance map, as shown in Extended data Fig.~\ref{Drange}. A unique trait of bilayer graphene is that the band gap size at charge neutrality reflects layer polarization. When the two layers become layer-degenerate, the system reaches a gapless condition. In Extended data Fig.~\ref{Drange}, the gapless points for the forward and backward scans are situated at positive and negative $D$ fields respectively. It means that, in the forward scan, an additional positive $D$ field is needed to compensate for a negative intrinsic polarization in order to reach the gapless condition and vice versa for the backward scan. The polarization change between the forward and backward scans can be calculated based on $\Delta D$ between the two gapless points, which is $\Delta P \approx 0.30 \mathrm{uC/cm^{-2}}$.

In addition, the electronic ratchet effect in regions \textbf{ii} and \textbf{iv} enables a continuously scalable ferroelectric response. The ferroelectric loop consists of regions \textbf{ii}, \textbf{iii}, and \textbf{iv}. While regime \textbf{iii} remains fixed in length, regions \textbf{ii} and \textbf{iv} can be elongated with increasing $D$ field range. The electronic ratchet effect described in Fig.~\ref{Fig3} ensures that reversing the scanning direction at any position in regimes \textbf{ii} and \textbf{iv} will trigger an immediate re-entrance of regime \textbf{iii}, hence binding the hysteresis loop. These unique properties result in a continuously scalable ferroelectric loop whose area depends linearly on $D$ field range, as illustrated by the black dashed line in Fig.~\ref{Fig3}\textbf{e}.

\subsection*{Programmable quasi-continuous memory states}

\begin{figure*}
    \includegraphics[width=4in]{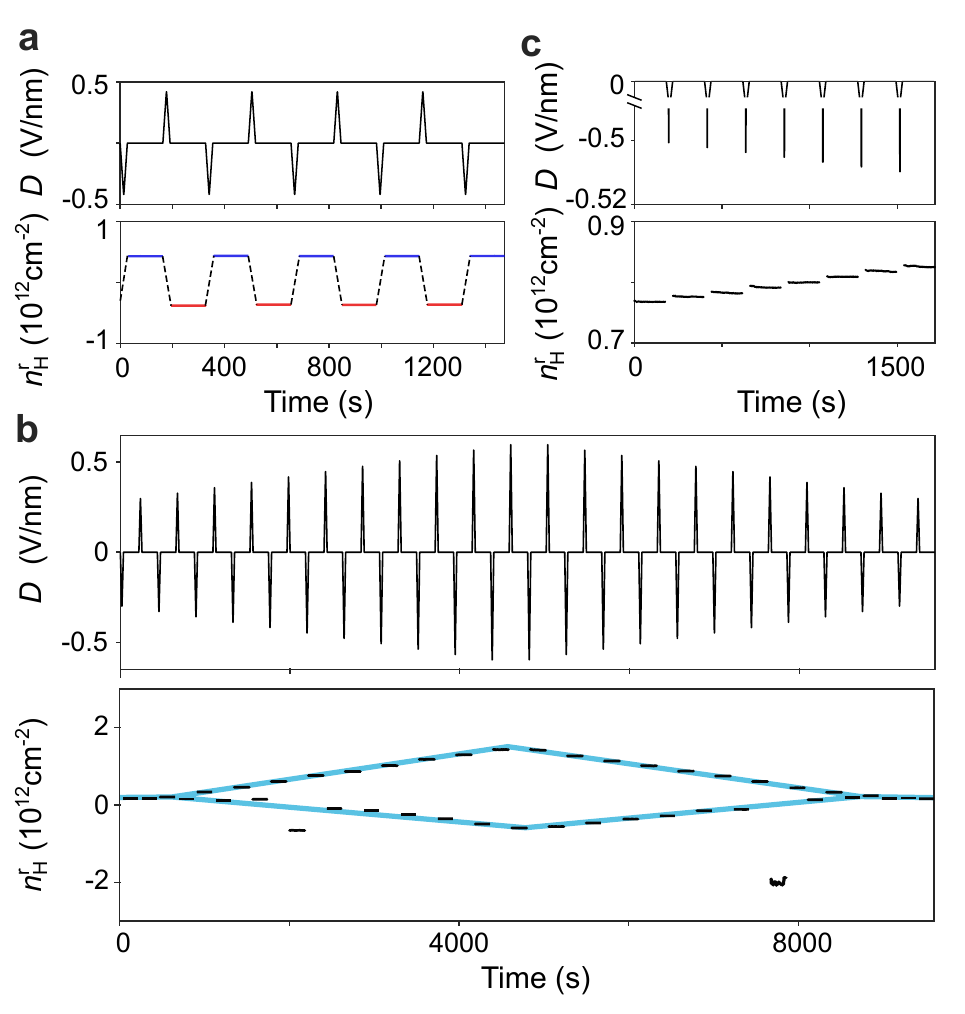}
    \caption{\figtitle{Programmable quasi-continuous memory states.} (\textbf{a}) At a fixed $n_\mathrm{Total}$, positive and negative $D$ field pulses are applied and the resulting $n_\mathrm{H}^\mathrm{r}$ at $D=0$ is measured as a function of time. $n_\mathrm{H}^\mathrm{r}$ shows a stable switching between positive and negative values, reflecting the stable switching between the up and down polarization states. For all the pulsing measurements plotted in the main text, only the remnant Hall density $n_\mathrm{H}^\mathrm{r}$ is shown, omitting the $n_\mathrm{H}$ during the pulses (See SI for full data). (\textbf{b}) A pulsing sequence with gradually increasing and decreasing magnitude is performed with the resulting $n_\mathrm{H}^\mathrm{r}$ plotted in the lower panel. Within a small $D$ field range, the system is non-hysteretic, where $n_\mathrm{H}^\mathrm{r}$ remains the same after positive and negative pulses. Beyond a critical $D$ field range, the system starts to develop remnant polarization. $n_\mathrm{H}^\mathrm{r}$ follows the lower and upper branches after the positive and negative pulses. As the magnitude of the pulses gradually decreases, $n_\mathrm{H}^\mathrm{r}$ responds accordingly and the system eventually returns to a non-hysteretic state. The two segments that deviates from the linear trend are due to the diverging behavior as the Hall density moves across the charge neutrality point. (\textbf{c}) A sequence of negative $D$ field pulses with small increments is applied with the resulting $n_\mathrm{H}^\mathrm{r}$ recorded. The $\Delta n$ as a result of negative pulses with small increments is estimated to be $6\times10^9\mathrm{cm}^{-2}$. With this resolution, our system can accommodate at least five hundred different memory states.}
    \label{Fig4_2}
\end{figure*}

The continuously scalable ferroelectric response can enable highly programmable memory states. In Fig.~\ref{Fig4_2}, we demonstrate this tunable functionality through pulse measurements in Device D2, which shares a similar behavior as Device D1 (Extended data Fig.~\ref{hyssummary}).

From the analysis above, we point out that the remnant itinerant charge density $n_\mathrm{H}^\mathrm{r}$ at $D=0$ is a direct consequence of the remnant polarization $P^\mathrm{r}$. At a fixed $n_\mathrm{Total}$, each $n_\mathrm{H}^\mathrm{r}$ value corresponds to a distinct memory state. In Fig.~\ref{Fig4_2}\textbf{a}, we repeatedly apply positive and negative $D$ field pulses to set the system to up and down polarized states and record $n_\mathrm{H}^\mathrm{r}$ as a function of time. The result demonstrates that the remnant polarization state remains stable over time when the external $D$ field is removed, indicating a non-volatile memory state (Extended data Fig.~\ref{Random}\textbf{a}).

Furthermore, by varying the pulses' magnitudes, we can deterministically program the system into distinct memory states, as shown in Fig.~\ref{Fig4_2}\textbf{b} (Extended data Fig.~\ref{Random}\textbf{b}). Within a small $D$ field range, the system behaves reversibly, namely $n_\mathrm{H}^\mathrm{r}$ remaining the same after the positive and negative pulses. Beyond a critical $D$ field range, the system starts to develop spontaneous polarization, leading to varying $n_\mathrm{H}^\mathrm{r}$. By increasing the pulse magnitude, we can induce a larger polarization and hence a greater value of $n_\mathrm{H}^\mathrm{r}$. But, surprisingly, reversing the pulse sequence, we can gradually decrease the magnitude of $n_\mathrm{H}^\mathrm{r}$ and revert the system from a hysteretic state to a non-hysteretic state.

These distinct memory states are distributed in a quasi-continuous fashion. By incrementing the $D$ field in tiny steps (Fig.~\ref{Fig4_2}\textbf{c}), we can reliably resolve distinct memory states down to the order of $n_\mathrm{H}^\mathrm{r} \approx 6\times10^{9}\mathrm{cm}^{-2}$. Hence, our system can accommodate at least $500$ different memory states (with total $n_\mathrm{H}^\mathrm{r}$ range $\approx 3\times10^{12}$). The resolution seems to be limited
only by the measurement noise level, device quality, angle disorder, and so on.

\subsection*{Indications of an electronic origin}

The above-described ferroelectric response is distinct from existing ferroelectrics and pose a challenge in its understanding under the framework of lattice-driven ferroelectrics. Below, we comment on its uniqueness.

The broken lattice inversion symmetry due to the layer-specific moir\'e potential is not sufficient to induce switchable dipole moments. From our discussion above, the layer-specific moir\'e potential is important in creating a contrasting electron wavefunction distribution in the two graphene layers, which lead to a top interface localized system and an itinerant system. However, if we assume that the explicit lattice inversion symmetry breaking generates a polar structure, achieving the opposite polarity would require a complete flipping of the entire structure. This is hardly achievable through electric field control due to the large angle difference between the two alignment schemes. Hence, the appearance of switchable dipole moments should involve mechanisms beyond the apparent lattice asymmetry.

In addition, our observed smooth ferroelectric switching is incompatible with sliding ferroelectricity. Recently, sliding ferroelectricity has become a powerful way to engineer 2D ferroelectricity, where polarization switching is achieved through a relative in-plane sliding of the constituent layers. However, since BLG is inversion symmetric, a structural analysis need to take into account both the top and bottom BN layers. For layer asymmetric alignment, calculations have shown that two different stacking configurations of BN-BLG-BN can possess two distinct vertical polarization in the same direction~\cite{yang2023across}. This is contrary to our quasi-continuous switchable states as shown above. Even with the consideration of domain pinning, the evolution of different polarization states in sliding ferroelectricity shows a step-like transition~\cite{Yasuda2021}, in direct contrast to the smooth ferroelectric switching curves, as shown in Fig.~\ref{Fig3}\textbf{e}. Hence, sliding ferroelectricity is unlikely to be the origin of our observed ferroelectric response.

Furthermore, the $D$ field dependence of polarization states, reflected by $n_\mathrm{H}^\mathrm{r}$, shows distinctive behavior from soft ferroelectrics. From the hysteresis loops shown in Fig.~\ref{Fig3}\textbf{e}, we can extract $n_\mathrm{H}^\mathrm{r}$ as a function of the maximum displacement field reached $D_\mathrm{max}$, as shown in Fig.~\ref{Fig4}\textbf{a}. The remnant $n_\mathrm{H}^\mathrm{r}$ shows a three-stage dependence on $D_\mathrm{max}$. Stage 1: when $2D_\mathrm{max}<w$ ($w$ is the $D$ range needed to traverse region \textbf{iii}, as labeled in Fig.~\ref{Fig3}\textbf{e} top loop), $n_\mathrm{H}^\mathrm{r}$ stays at zero. Stage 2: when $w<2D_\mathrm{max}<2w$, $n_\mathrm{H}^\mathrm{r}$ grows proportionally to $D_\mathrm{max}$. Stage 3: when $2D_\mathrm{max}>2w$, $n_\mathrm{H}^\mathrm{r}$ reaches saturation. To the best of our knowledge, the continuous and linear tuning of $n_\mathrm{H}^\mathrm{r}$, and subsequently, $P^\mathrm{r}$, by $D_\mathrm{max}$ and ever-growing rigid loops with $D_\mathrm{max}$ ($w<2D_\mathrm{max}<2w$) are distinct from existing ferroelectrics~\cite{gonzalo2005ferroelectricity}, including soft ferroelectrics.

Altogether, our observation of continuously scalable ferroelectric response enabled by the electronic ratchet effect displays unique properties that cannot be understood solely based on lattice-driven mechanisms. It strongly indicates that the spontaneous polarization in our system emerges from an electronic origin, which calls for a new theoretical framework for its understanding.

\subsection*{A phenomenological model based on ferroelectric units}

\begin{figure*}
    \includegraphics[width=6.5in]{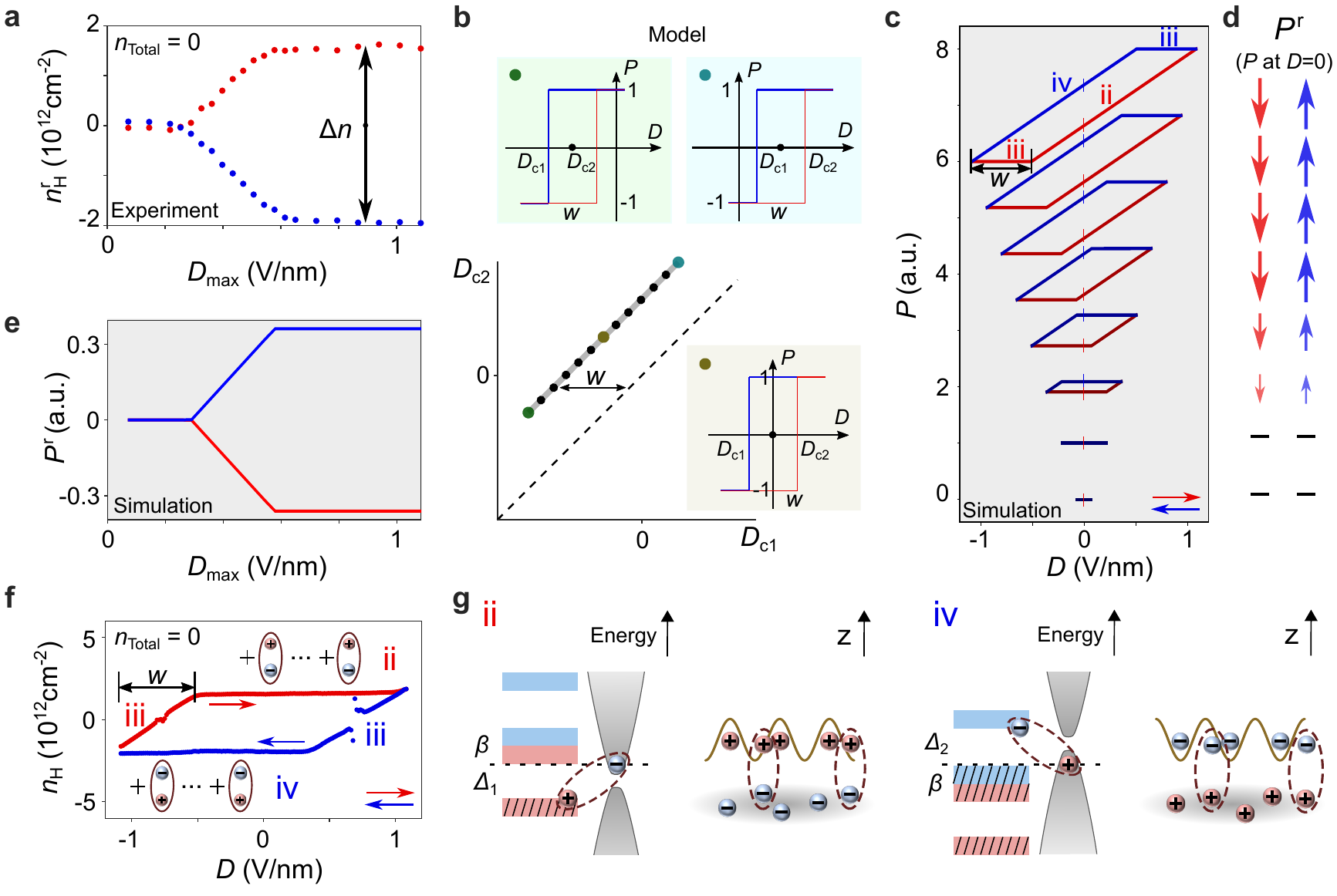}
    \caption{\figtitle{Simulation of scalable ferroelectricity by a phenomenological model and a picture of excitonic ferroelectricity.} (\textbf{a}) $n_\mathrm{H}^\mathrm{r}$ measured at $n_\mathrm{Total}=0$ as a function of $D_\mathrm{max}$. When $2D_\mathrm{max}>w$, $n_\mathrm{H}^\mathrm{r}$ becomes non-zero and shows a linear dependence on $D_\mathrm{max}$ before saturation. (\textbf{b}) A phenomenological model with a unique set of ferroelectric units called hysterons. Each hysteron is characterized by its critical electric displacement fields, $D_\mathrm{c1}$ and $D_\mathrm{c2}$. To simulate our system, we have an evenly distributed hysteron along a finite segment within the $(D_\mathrm{c1}, D_\mathrm{c2})$ phase space. The three example insets correspond to the colored dots in the main panel. (\textbf{c}) The simulated polarization $P$ plotted as a function of $D$ for different $D$ ranges. The constant polarization segment has a fixed width $w$. (\textbf{d}) The relative scale of remnant polarization $P^\mathrm{r}$ is illustrated for each $D$ range. (\textbf{e}) The simulated $P^\mathrm{r}$ plotted as a function of $D_\mathrm{max}$, showing a three-stage behavior that is consistent with $n_\mathrm{H}^\mathrm{r}$. (\textbf{f}) A picture of dipolar excitons as the hysterons in our system. Excitons are generated in regions \textbf{ii} and \textbf{iv} with opposite polarization. The length of the LSAS regime, which corresponds to the $D$ field needed to tune the filling of the localized system in between $n_\mathrm{L}=\pm n_0$, defines the width $w$ in the hysteron model. (\textbf{g}) Schematics of exciton formation. Within region \textbf{ii} ($\Delta_1$), the newly added holes in the localized subsystem would experience a large Coulomb repulsion. Based on our calculation of the exciton binding energy, it is favorable for these holes in the localized subsystem and electrons in the itinerant subsystem to form exciton pairs, lowering the overall energy of the system. Within region \textbf{iv} ($\Delta_2$), the process mirrored the same behavior in region \textbf{ii} by replacing holes with electrons, resulting in bound pairs of localized electrons and itinerant holes.
    }
    \label{Fig4}
\end{figure*}

Below, we present a phenomenological model to simulate the ferroelectric response of our system, which successfully captures the above unique observations. As it will be shown, such agreement strongly suggests that our system has a unique set of ferroelectric units. Our approach is inspired by the Preisach model proposed in 1935~\cite{preisach1935magnetische} to describe hysteretic phenomena, including ferromagnetism and ferroelectricity. This model describes ferroelectricity by ferroelectric units, the so-called hysterons. Each hysteron is characterized by a rectangular hysteresis loop, similar to the ones shown in Fig.~\ref{Fig4}\textbf{b}, with critical displacements fields $D_\mathrm{c1}$ and $D_\mathrm{c2}$. $D_\mathrm{c1}$ and $D_\mathrm{c2}$ can vary for each hysteron and together form a parameter space. The system's total polarization $P$ at any given $D$ field is the sum of the individual polarizations of each hysteron. Hence, the entire ferroelectric response, namely the $P-D$ loop, is the sum of individual $P-D$ loops of hysterons. For example, ideal hard ferroelectrics consist of hysterons with the same $D_\mathrm{c1}$ and $D_\mathrm{c2}$. Soft ferroelectrics consist of hysterons whose $D_\mathrm{c1}$ and $D_\mathrm{c2}$ independently follow a Gaussian distribution~\cite{cima2002characterization}.

In order to simulate our ferroelectric response, we found that the hysteron distribution is strongly constrained to the following characteristics: \textbf{(1)} The rectangular hysteresis loop has a fixed width, i.e., $D_\mathrm{c2}-D_\mathrm{c1}=w$ ($D$ is the field directly controlled in our experiment).  \textbf{(2)} The center of the loop needs to be densely spaced and uniformly distributed along the $D$ axis. \textbf{(3)} Hysterons exist for $D_\mathrm{c2} > D_\mathrm{c1} > 0$ and $0 > D_\mathrm{c2} > D_\mathrm{c1}$. The conditions \textbf{(1)}-\textbf{(3)} translate into Fig.~\ref{Fig4}\textbf{b}, where $D_\mathrm{c1}$ and $D_\mathrm{c2}$ are uniformly and densely distributed along a line ($D_\mathrm{c2}-D_\mathrm{c1}=w$) extending to both $(+, +)$ and $(-, -)$ quadrants.

Our simulated results in Figs.~\ref{Fig4}\textbf{c-e} capture the key observations in our experiments (Fig.~\ref{Fig3}\textbf{e} and Fig.~\ref{Fig4}\textbf{a}), including the continuously scalable hysteresis loops and the three-stage evolution of $P^\mathrm{r}$ with respect to $D_\mathrm{max}$ (see the comprehensive discussion in SI.D.1). The one-to-one correspondence between $P$ and $n_\mathrm{H}$ is of particular importance as it reveals that the ratcheting regions \textbf{ii} and \textbf{iv} play a crucial role in driving the polarization change. This relationship can be directly visualized through a capacitor model detailed in SI.D.3.

\subsection*{Dipolar excitons as ferroelectric units}

Our discussion above demonstrates the uniqueness of the continuously scalable ferroelectric response, which is likely of an electronic origin and can be simulated by a set of versatile ferroelectric units. One particularly promising candidate for an electronic ferroelectric unit is a dipolar exciton. Such a quantum object consists of an electron and a hole that are spatially separated and it carries the smallest unit of electric dipole moment. Below, we explain the formation process of excitons and discuss how excitons can account for our main experimental observations and satisfy the three characteristics \textbf{(1)} to \textbf{(3)} of hysterons.

Excitons are formed in regions \textbf{ii} and \textbf{iv}, shown in Fig.~\ref{Fig4}\textbf{f}. Due to the electron-hole symmetry, we focus on region \textbf{ii} where band $\beta$ is at filling of one hole per flavor in each moir\'e site ($n_\mathrm{L} = -n_0$). Further decreasing $V_\mathrm{TG}$ tends to polarize (add) more holes to $n_\mathrm{L}$. However, an excess hole in the localized subsystem would trigger double occupancy with a cost of Coulomb energy $U$. Alternatively, this hole in $n_\mathrm{L}$ can bind with an electron in $n_\mathrm{H}$ to form an exciton, lowering the energy by $E_\mathrm{B}$ (exciton binding energy). Hence, the varying $P$ in the ratcheting regimes can be understood as the formation of excitons with upward and downward polarities in region \textbf{ii} and \textbf{iv} (Fig.~\ref{Fig4}\textbf{g}).

The formation of excitons can naturally explain our main experimental observations. There are two parts. First, $V_\mathrm{TG}$ can induce free carriers in the itinerant system in the ratcheting regimes. In region \textbf{ii}, $V_\mathrm{TG}$ adds $N$ holes to the localized system in the form of $N$ excitons, by binding $N$ electrons from the itinerant system. From charge conservation, an equal amount of mobile holes needs to be added to the itinerant system. Therefore, by lowering $V_\mathrm{TG}$ in region \textbf{ii}, we observe the decrease of $n_\mathrm{H}$ (an increase of hole density), as shown in Fig.~\ref{Fig3}\textbf{c}. If we consider the $D$ field sweep, the change of $n_\mathrm{H}$ will be compensated by $V_\mathrm{BG}$, as shown in Fig.~\ref{Fig3}\textbf{e}.
Second, the formation of exciton explains the ``locking'' mechanism. Once the exciton is formed, the system, including the exciton, localized and itinerant charges, effectively enters a many-body phase due to interaction effects.
Reversing the process does not immediately break the excitons. Furthermore, this exciton generates an internal electric field that compensates the external electric field by $V_\mathrm{TG}$ in region \textbf{ii}. As a result, the Fermi level is pinned to the bottom of the localized band $\beta$ in region \textbf{ii}. Reversing the scan of $V_\mathrm{TG}$ immediately triggers the re-entrance into region \textbf{iii}.

With the above description, we can review the characteristics of hysterons in our model within the exciton picture. \textbf{(1)} The fixed width of the individual hysteron loop, $w$, corresponds to the fixed $D$ range (or $V_\mathrm{TG}$ range) needed to initiate the formation of oppositely polarized excitons, during which $n_\mathrm{L}$ changes between $-n_0$ and $+n_0$. \textbf{(2)} The dense and uniform hysteron distribution is reflected in the continuous variation of exciton density, and hence polarization, that can be linearly controlled by $V_\mathrm{TG}$. \textbf{(3)} The onset of exciton formation and its polarity are determined by the filling of the localized subsystem, instead of the external $D$. Therefore, the system can form excitons opposite to the direction of $D$, leading to hysterons with $D_\mathrm{c2} > D_\mathrm{c1} > 0$ and $0 > D_\mathrm{c2} > D_\mathrm{c1}$.

Taken together, dipolar excitons stand as the most natural candidate for the electronic ferroelectric units and the formation of excitons actually drives the unique ferroelectric response in our system. We further calculate the exciton binding energy $E_\mathrm{B}$ between a top-layer polarized flat band and a BLG-like dispersive band, which is on the order of 100 meV (see SI.E.1). This value is comparable to the on-site $U$ extracted from the experiment. It, indeed, suggests favorable conditions for exciton formation in our system.

\subsection*{Outlook}
Our experimental findings, in conjunction with our simulations and theoretical analysis, provide the first experimental signatures of excitonic ferroelectricity. Although we present a phenomenological theory to explain this remarkable observation, many unanswered questions remain, which will require further experimental and theoretical investigation. For instance, while we currently believe that the alignment angle for the itinerant side is not critical, as long as it is sufficiently large, it is possible that special combinations of angles exist (in terms of alignment with top and bottom BN layers), which can achieve the dichotomy of localized and itinerant subsystems while maintaining strong Coulomb coupling between them. We note in that sense that the devices where we have observed the excitonic ferroelectricity phenomenology had $\sim$ 15$\degree$ or 30$\degree$ on the itinerant side. However, a detailed study of the angular dependence~\cite{ribeiro2018twistable, inbar2023quantum} of dual graphene-BN moir\'e structures with well-defined angles, a substantial challenge itself, would be highly desirable. In addition, while we have obtained direct evidence for itinerant and localized charges through our transport measurement (Hall density and Landau level measurement, see SI), it would be beneficial to directly image the localized charges using spectroscopy. It is worth noting that in our device, the distinct gate tunability necessitates the use of a top gate to adjust the Fermi level of the localized system, which introduces the need for specialized measurement schemes~\cite{li2021imaging}.

Moreover, understanding how charge density and temperature affect the binding and screening of excitons is an ongoing question in our study. In this respect, we briefly describe two additional observations that could provide valuable information. First, we have observed that the ratcheting behavior exists for a wide range of $V_\mathrm{BG}$ values that tune the density of mobile charge, but it can be suppressed when $V_\mathrm{BG}$ exceeds a certain threshold value (see SI.E.2). The observed effect may be due to the influence of $V_\mathrm{BG}$ on the relative energy shift between the localized and itinerant electronic states, which affects exciton formation. Second, the observed ferroelectric response remains resilient to high temperatures and can be present even at room temperature~\cite{Yan2023The}. The robustness of exciton formation in our system is unsurprising, given the large Coulomb energy scale. However, exciton condensation is less likely to occur at high temperatures, suggesting that dipolar excitons at finite density may maintain the ferroelectric response, in a different regime from the POS model \cite{portengen1996theory}. Given the similarities
, it is intriguing to investigate the relevance of the POS model in our study and to investigate whether new phenomena can occur below a critical temperature. We also note that although our experimental data is consistent with electronically driven ferroelectricity, one cannot completely disentangle the electronic and lattice degrees of freedom. It is therefore an interesting future direction to investigate the couplings of excitons with structural changes, e.g., through electronic polarons.

We also emphasize that our findings could have significant implications beyond our specific moir\'e system. Specifically, our interpretation of the unconventional hysteretic phenomena observed in our study could serve as a foundation for understanding related unusual ferroelectric-like responses observed in other systems~\cite{klein2023electrical, niu2022giant}, and provide guidelines for designing new 2D electronic ferroelectrics and even multiferroic systems. Furthermore, the highly unusual ratcheting charge dynamics could lead to the development of next-generation memory devices and synaptic transistors used in neuromorphic computing~\cite{christensen20222022, Yan2023The}. Moreover, the electronic origin of the ferroelectricity suggests ultrafast dynamics, whose study will require time-resolved experiments.

\setcitestyle{numbers}
\bibliography{references}

\begin{thebibliography}{48}%
\makeatletter
\providecommand \@ifxundefined [1]{%
 \@ifx{#1\undefined}
}%
\providecommand \@ifnum [1]{%
 \ifnum #1\expandafter \@firstoftwo
 \else \expandafter \@secondoftwo
 \fi
}%
\providecommand \@ifx [1]{%
 \ifx #1\expandafter \@firstoftwo
 \else \expandafter \@secondoftwo
 \fi
}%
\providecommand \natexlab [1]{#1}%
\providecommand \enquote  [1]{``#1''}%
\providecommand \bibnamefont  [1]{#1}%
\providecommand \bibfnamefont [1]{#1}%
\providecommand \citenamefont [1]{#1}%
\providecommand \href@noop [0]{\@secondoftwo}%
\providecommand \href [0]{\begingroup \@sanitize@url \@href}%
\providecommand \@href[1]{\@@startlink{#1}\@@href}%
\providecommand \@@href[1]{\endgroup#1\@@endlink}%
\providecommand \@sanitize@url [0]{\catcode `\\12\catcode `\$12\catcode
  `\&12\catcode `\#12\catcode `\^12\catcode `\_12\catcode `\%12\relax}%
\providecommand \@@startlink[1]{}%
\providecommand \@@endlink[0]{}%
\providecommand \url  [0]{\begingroup\@sanitize@url \@url }%
\providecommand \@url [1]{\endgroup\@href {#1}{\urlprefix }}%
\providecommand \urlprefix  [0]{URL }%
\providecommand \Eprint [0]{\href }%
\providecommand \doibase [0]{https://doi.org/}%
\providecommand \selectlanguage [0]{\@gobble}%
\providecommand \bibinfo  [0]{\@secondoftwo}%
\providecommand \bibfield  [0]{\@secondoftwo}%
\providecommand \translation [1]{[#1]}%
\providecommand \BibitemOpen [0]{}%
\providecommand \bibitemStop [0]{}%
\providecommand \bibitemNoStop [0]{.\EOS\space}%
\providecommand \EOS [0]{\spacefactor3000\relax}%
\providecommand \BibitemShut  [1]{\csname bibitem#1\endcsname}%
\let\auto@bib@innerbib\@empty
\bibitem [{\citenamefont {Ishihara}(2010)}]{ishihara2010electronic}%
  \BibitemOpen
  \bibfield  {author} {\bibinfo {author} {\bibfnamefont {S.}~\bibnamefont
  {Ishihara}},\ }\bibfield  {title} {\bibinfo {title} {Electronic
  ferroelectricity and frustration},\ }\href@noop {} {\bibfield  {journal}
  {\bibinfo  {journal} {Journal of the Physical Society of Japan}\ }\textbf
  {\bibinfo {volume} {79}},\ \bibinfo {pages} {011010} (\bibinfo {year}
  {2010})}\BibitemShut {NoStop}%
\bibitem [{\citenamefont {Yamauchi}\ and\ \citenamefont
  {Barone}(2014)}]{yamauchi2014electronic}%
  \BibitemOpen
  \bibfield  {author} {\bibinfo {author} {\bibfnamefont {K.}~\bibnamefont
  {Yamauchi}}\ and\ \bibinfo {author} {\bibfnamefont {P.}~\bibnamefont
  {Barone}},\ }\bibfield  {title} {\bibinfo {title} {Electronic
  ferroelectricity induced by charge and orbital orderings},\ }\href@noop {}
  {\bibfield  {journal} {\bibinfo  {journal} {Journal of Physics: Condensed
  Matter}\ }\textbf {\bibinfo {volume} {26}},\ \bibinfo {pages} {103201}
  (\bibinfo {year} {2014})}\BibitemShut {NoStop}%
\bibitem [{\citenamefont {Tokura}\ \emph {et~al.}(2017)\citenamefont {Tokura},
  \citenamefont {Kawasaki},\ and\ \citenamefont
  {Nagaosa}}]{tokura2017emergent}%
  \BibitemOpen
  \bibfield  {author} {\bibinfo {author} {\bibfnamefont {Y.}~\bibnamefont
  {Tokura}}, \bibinfo {author} {\bibfnamefont {M.}~\bibnamefont {Kawasaki}},\
  and\ \bibinfo {author} {\bibfnamefont {N.}~\bibnamefont {Nagaosa}},\
  }\bibfield  {title} {\bibinfo {title} {Emergent functions of quantum
  materials},\ }\href@noop {} {\bibfield  {journal} {\bibinfo  {journal}
  {Nature Physics}\ }\textbf {\bibinfo {volume} {13}},\ \bibinfo {pages} {1056}
  (\bibinfo {year} {2017})}\BibitemShut {NoStop}%
\bibitem [{\citenamefont {Portengen}\ \emph {et~al.}(1996)\citenamefont
  {Portengen}, \citenamefont {{\"O}streich},\ and\ \citenamefont
  {Sham}}]{portengen1996theory}%
  \BibitemOpen
  \bibfield  {author} {\bibinfo {author} {\bibfnamefont {T.}~\bibnamefont
  {Portengen}}, \bibinfo {author} {\bibfnamefont {T.}~\bibnamefont
  {{\"O}streich}},\ and\ \bibinfo {author} {\bibfnamefont {L.}~\bibnamefont
  {Sham}},\ }\bibfield  {title} {\bibinfo {title} {Theory of electronic
  ferroelectricity},\ }\href@noop {} {\bibfield  {journal} {\bibinfo  {journal}
  {Physical Review B}\ }\textbf {\bibinfo {volume} {54}},\ \bibinfo {pages}
  {17452} (\bibinfo {year} {1996})}\BibitemShut {NoStop}%
\bibitem [{\citenamefont {Batyev}\ and\ \citenamefont
  {Borisyuk}(1980)}]{batyev1980excitonic}%
  \BibitemOpen
  \bibfield  {author} {\bibinfo {author} {\bibfnamefont {{\'E}.}~\bibnamefont
  {Batyev}}\ and\ \bibinfo {author} {\bibfnamefont {V.}~\bibnamefont
  {Borisyuk}},\ }\bibfield  {title} {\bibinfo {title} {Excitonic insulator as a
  ferroelectric material},\ }\href@noop {} {\bibfield  {journal} {\bibinfo
  {journal} {Soviet Journal of Experimental and Theoretical Physics Letters}\
  }\textbf {\bibinfo {volume} {32}},\ \bibinfo {pages} {395} (\bibinfo {year}
  {1980})}\BibitemShut {NoStop}%
\bibitem [{\citenamefont {Batista}(2002)}]{batista2002electronic}%
  \BibitemOpen
  \bibfield  {author} {\bibinfo {author} {\bibfnamefont {C.}~\bibnamefont
  {Batista}},\ }\bibfield  {title} {\bibinfo {title} {Electronic
  ferroelectricity in the falicov-kimball model},\ }\href@noop {} {\bibfield
  {journal} {\bibinfo  {journal} {Physical review letters}\ }\textbf {\bibinfo
  {volume} {89}},\ \bibinfo {pages} {166403} (\bibinfo {year}
  {2002})}\BibitemShut {NoStop}%
\bibitem [{\citenamefont {Nandkishore}\ and\ \citenamefont
  {Levitov}(2010)}]{nandkishore2010dynamical}%
  \BibitemOpen
  \bibfield  {author} {\bibinfo {author} {\bibfnamefont {R.}~\bibnamefont
  {Nandkishore}}\ and\ \bibinfo {author} {\bibfnamefont {L.}~\bibnamefont
  {Levitov}},\ }\bibfield  {title} {\bibinfo {title} {Dynamical screening and
  excitonic instability in bilayer graphene},\ }\href@noop {} {\bibfield
  {journal} {\bibinfo  {journal} {Physical review letters}\ }\textbf {\bibinfo
  {volume} {104}},\ \bibinfo {pages} {156803} (\bibinfo {year}
  {2010})}\BibitemShut {NoStop}%
\bibitem [{\citenamefont {Christensen}\ \emph {et~al.}(2022)\citenamefont
  {Christensen}, \citenamefont {Dittmann}, \citenamefont {Linares-Barranco},
  \citenamefont {Sebastian}, \citenamefont {Le~Gallo}, \citenamefont
  {Redaelli}, \citenamefont {Slesazeck}, \citenamefont {Mikolajick},
  \citenamefont {Spiga}, \citenamefont {Menzel} \emph
  {et~al.}}]{christensen20222022}%
  \BibitemOpen
  \bibfield  {author} {\bibinfo {author} {\bibfnamefont {D.~V.}\ \bibnamefont
  {Christensen}}, \bibinfo {author} {\bibfnamefont {R.}~\bibnamefont
  {Dittmann}}, \bibinfo {author} {\bibfnamefont {B.}~\bibnamefont
  {Linares-Barranco}}, \bibinfo {author} {\bibfnamefont {A.}~\bibnamefont
  {Sebastian}}, \bibinfo {author} {\bibfnamefont {M.}~\bibnamefont {Le~Gallo}},
  \bibinfo {author} {\bibfnamefont {A.}~\bibnamefont {Redaelli}}, \bibinfo
  {author} {\bibfnamefont {S.}~\bibnamefont {Slesazeck}}, \bibinfo {author}
  {\bibfnamefont {T.}~\bibnamefont {Mikolajick}}, \bibinfo {author}
  {\bibfnamefont {S.}~\bibnamefont {Spiga}}, \bibinfo {author} {\bibfnamefont
  {S.}~\bibnamefont {Menzel}}, \emph {et~al.},\ }\bibfield  {title} {\bibinfo
  {title} {2022 roadmap on neuromorphic computing and engineering},\
  }\href@noop {} {\bibfield  {journal} {\bibinfo  {journal} {Neuromorphic
  Computing and Engineering}\ }\textbf {\bibinfo {volume} {2}},\ \bibinfo
  {pages} {022501} (\bibinfo {year} {2022})}\BibitemShut {NoStop}%
\bibitem [{\citenamefont {Cao}\ \emph {et~al.}(2018{\natexlab{a}})\citenamefont
  {Cao}, \citenamefont {Fatemi}, \citenamefont {Demir}, \citenamefont {Fang},
  \citenamefont {Tomarken}, \citenamefont {Luo}, \citenamefont
  {Sanchez-Yamagishi}, \citenamefont {Watanabe}, \citenamefont {Taniguchi},
  \citenamefont {Kaxiras} \emph {et~al.}}]{cao2018correlated}%
  \BibitemOpen
  \bibfield  {author} {\bibinfo {author} {\bibfnamefont {Y.}~\bibnamefont
  {Cao}}, \bibinfo {author} {\bibfnamefont {V.}~\bibnamefont {Fatemi}},
  \bibinfo {author} {\bibfnamefont {A.}~\bibnamefont {Demir}}, \bibinfo
  {author} {\bibfnamefont {S.}~\bibnamefont {Fang}}, \bibinfo {author}
  {\bibfnamefont {S.~L.}\ \bibnamefont {Tomarken}}, \bibinfo {author}
  {\bibfnamefont {J.~Y.}\ \bibnamefont {Luo}}, \bibinfo {author} {\bibfnamefont
  {J.~D.}\ \bibnamefont {Sanchez-Yamagishi}}, \bibinfo {author} {\bibfnamefont
  {K.}~\bibnamefont {Watanabe}}, \bibinfo {author} {\bibfnamefont
  {T.}~\bibnamefont {Taniguchi}}, \bibinfo {author} {\bibfnamefont
  {E.}~\bibnamefont {Kaxiras}}, \emph {et~al.},\ }\bibfield  {title} {\bibinfo
  {title} {Correlated insulator behaviour at half-filling in magic-angle
  graphene superlattices},\ }\href@noop {} {\bibfield  {journal} {\bibinfo
  {journal} {Nature}\ }\textbf {\bibinfo {volume} {556}},\ \bibinfo {pages}
  {80} (\bibinfo {year} {2018}{\natexlab{a}})}\BibitemShut {NoStop}%
\bibitem [{\citenamefont {Cao}\ \emph {et~al.}(2018{\natexlab{b}})\citenamefont
  {Cao}, \citenamefont {Fatemi}, \citenamefont {Fang}, \citenamefont
  {Watanabe}, \citenamefont {Taniguchi}, \citenamefont {Kaxiras},\ and\
  \citenamefont {Jarillo-Herrero}}]{cao2018unconventional}%
  \BibitemOpen
  \bibfield  {author} {\bibinfo {author} {\bibfnamefont {Y.}~\bibnamefont
  {Cao}}, \bibinfo {author} {\bibfnamefont {V.}~\bibnamefont {Fatemi}},
  \bibinfo {author} {\bibfnamefont {S.}~\bibnamefont {Fang}}, \bibinfo {author}
  {\bibfnamefont {K.}~\bibnamefont {Watanabe}}, \bibinfo {author}
  {\bibfnamefont {T.}~\bibnamefont {Taniguchi}}, \bibinfo {author}
  {\bibfnamefont {E.}~\bibnamefont {Kaxiras}},\ and\ \bibinfo {author}
  {\bibfnamefont {P.}~\bibnamefont {Jarillo-Herrero}},\ }\bibfield  {title}
  {\bibinfo {title} {Unconventional superconductivity in magic-angle graphene
  superlattices},\ }\href@noop {} {\bibfield  {journal} {\bibinfo  {journal}
  {Nature}\ }\textbf {\bibinfo {volume} {556}},\ \bibinfo {pages} {43}
  (\bibinfo {year} {2018}{\natexlab{b}})}\BibitemShut {NoStop}%
\bibitem [{\citenamefont {Andrei}\ and\ \citenamefont
  {MacDonald}(2020)}]{andrei2020graphene}%
  \BibitemOpen
  \bibfield  {author} {\bibinfo {author} {\bibfnamefont {E.~Y.}\ \bibnamefont
  {Andrei}}\ and\ \bibinfo {author} {\bibfnamefont {A.~H.}\ \bibnamefont
  {MacDonald}},\ }\bibfield  {title} {\bibinfo {title} {Graphene bilayers with
  a twist},\ }\href@noop {} {\bibfield  {journal} {\bibinfo  {journal} {Nature
  Materials}\ }\textbf {\bibinfo {volume} {19}},\ \bibinfo {pages} {1265}
  (\bibinfo {year} {2020})}\BibitemShut {NoStop}%
\bibitem [{\citenamefont {Andrei}\ \emph {et~al.}(2021)\citenamefont {Andrei},
  \citenamefont {Efetov}, \citenamefont {Jarillo-Herrero}, \citenamefont
  {MacDonald}, \citenamefont {Mak}, \citenamefont {Senthil}, \citenamefont
  {Tutuc}, \citenamefont {Yazdani},\ and\ \citenamefont
  {Young}}]{andrei2021marvels}%
  \BibitemOpen
  \bibfield  {author} {\bibinfo {author} {\bibfnamefont {E.~Y.}\ \bibnamefont
  {Andrei}}, \bibinfo {author} {\bibfnamefont {D.~K.}\ \bibnamefont {Efetov}},
  \bibinfo {author} {\bibfnamefont {P.}~\bibnamefont {Jarillo-Herrero}},
  \bibinfo {author} {\bibfnamefont {A.~H.}\ \bibnamefont {MacDonald}}, \bibinfo
  {author} {\bibfnamefont {K.~F.}\ \bibnamefont {Mak}}, \bibinfo {author}
  {\bibfnamefont {T.}~\bibnamefont {Senthil}}, \bibinfo {author} {\bibfnamefont
  {E.}~\bibnamefont {Tutuc}}, \bibinfo {author} {\bibfnamefont
  {A.}~\bibnamefont {Yazdani}},\ and\ \bibinfo {author} {\bibfnamefont {A.~F.}\
  \bibnamefont {Young}},\ }\bibfield  {title} {\bibinfo {title} {The marvels of
  moir{\'e} materials},\ }\href@noop {} {\bibfield  {journal} {\bibinfo
  {journal} {Nature Reviews Materials}\ }\textbf {\bibinfo {volume} {6}},\
  \bibinfo {pages} {201} (\bibinfo {year} {2021})}\BibitemShut {NoStop}%
\bibitem [{\citenamefont {Mak}\ and\ \citenamefont
  {Shan}(2022)}]{mak2022semiconductor}%
  \BibitemOpen
  \bibfield  {author} {\bibinfo {author} {\bibfnamefont {K.~F.}\ \bibnamefont
  {Mak}}\ and\ \bibinfo {author} {\bibfnamefont {J.}~\bibnamefont {Shan}},\
  }\bibfield  {title} {\bibinfo {title} {Semiconductor moir{\'e} materials},\
  }\href@noop {} {\bibfield  {journal} {\bibinfo  {journal} {Nature
  Nanotechnology}\ }\textbf {\bibinfo {volume} {17}},\ \bibinfo {pages} {686}
  (\bibinfo {year} {2022})}\BibitemShut {NoStop}%
\bibitem [{\citenamefont {Rickhaus}\ \emph {et~al.}(2021)\citenamefont
  {Rickhaus}, \citenamefont {de~Vries}, \citenamefont {Zhu}, \citenamefont
  {Portoles}, \citenamefont {Zheng}, \citenamefont {Masseroni}, \citenamefont
  {Kurzmann}, \citenamefont {Taniguchi}, \citenamefont {Watanabe},
  \citenamefont {MacDonald} \emph {et~al.}}]{rickhaus2021correlated}%
  \BibitemOpen
  \bibfield  {author} {\bibinfo {author} {\bibfnamefont {P.}~\bibnamefont
  {Rickhaus}}, \bibinfo {author} {\bibfnamefont {F.~K.}\ \bibnamefont
  {de~Vries}}, \bibinfo {author} {\bibfnamefont {J.}~\bibnamefont {Zhu}},
  \bibinfo {author} {\bibfnamefont {E.}~\bibnamefont {Portoles}}, \bibinfo
  {author} {\bibfnamefont {G.}~\bibnamefont {Zheng}}, \bibinfo {author}
  {\bibfnamefont {M.}~\bibnamefont {Masseroni}}, \bibinfo {author}
  {\bibfnamefont {A.}~\bibnamefont {Kurzmann}}, \bibinfo {author}
  {\bibfnamefont {T.}~\bibnamefont {Taniguchi}}, \bibinfo {author}
  {\bibfnamefont {K.}~\bibnamefont {Watanabe}}, \bibinfo {author}
  {\bibfnamefont {A.~H.}\ \bibnamefont {MacDonald}}, \emph {et~al.},\
  }\bibfield  {title} {\bibinfo {title} {Correlated electron-hole state in
  twisted double-bilayer graphene},\ }\href@noop {} {\bibfield  {journal}
  {\bibinfo  {journal} {Science}\ }\textbf {\bibinfo {volume} {373}},\ \bibinfo
  {pages} {1257} (\bibinfo {year} {2021})}\BibitemShut {NoStop}%
\bibitem [{\citenamefont {Gu}\ \emph {et~al.}(2022)\citenamefont {Gu},
  \citenamefont {Ma}, \citenamefont {Liu}, \citenamefont {Watanabe},
  \citenamefont {Taniguchi}, \citenamefont {Hone}, \citenamefont {Shan},\ and\
  \citenamefont {Mak}}]{gu2022dipolar}%
  \BibitemOpen
  \bibfield  {author} {\bibinfo {author} {\bibfnamefont {J.}~\bibnamefont
  {Gu}}, \bibinfo {author} {\bibfnamefont {L.}~\bibnamefont {Ma}}, \bibinfo
  {author} {\bibfnamefont {S.}~\bibnamefont {Liu}}, \bibinfo {author}
  {\bibfnamefont {K.}~\bibnamefont {Watanabe}}, \bibinfo {author}
  {\bibfnamefont {T.}~\bibnamefont {Taniguchi}}, \bibinfo {author}
  {\bibfnamefont {J.~C.}\ \bibnamefont {Hone}}, \bibinfo {author}
  {\bibfnamefont {J.}~\bibnamefont {Shan}},\ and\ \bibinfo {author}
  {\bibfnamefont {K.~F.}\ \bibnamefont {Mak}},\ }\bibfield  {title} {\bibinfo
  {title} {Dipolar excitonic insulator in a moire lattice},\ }\href@noop {}
  {\bibfield  {journal} {\bibinfo  {journal} {Nature Physics}\ }\textbf
  {\bibinfo {volume} {18}},\ \bibinfo {pages} {395} (\bibinfo {year}
  {2022})}\BibitemShut {NoStop}%
\bibitem [{\citenamefont {Chen}\ \emph {et~al.}(2022)\citenamefont {Chen},
  \citenamefont {Lian}, \citenamefont {Huang}, \citenamefont {Su},
  \citenamefont {Rashetnia}, \citenamefont {Ma}, \citenamefont {Yan},
  \citenamefont {Blei}, \citenamefont {Xiang}, \citenamefont {Taniguchi} \emph
  {et~al.}}]{chen2022excitonic}%
  \BibitemOpen
  \bibfield  {author} {\bibinfo {author} {\bibfnamefont {D.}~\bibnamefont
  {Chen}}, \bibinfo {author} {\bibfnamefont {Z.}~\bibnamefont {Lian}}, \bibinfo
  {author} {\bibfnamefont {X.}~\bibnamefont {Huang}}, \bibinfo {author}
  {\bibfnamefont {Y.}~\bibnamefont {Su}}, \bibinfo {author} {\bibfnamefont
  {M.}~\bibnamefont {Rashetnia}}, \bibinfo {author} {\bibfnamefont
  {L.}~\bibnamefont {Ma}}, \bibinfo {author} {\bibfnamefont {L.}~\bibnamefont
  {Yan}}, \bibinfo {author} {\bibfnamefont {M.}~\bibnamefont {Blei}}, \bibinfo
  {author} {\bibfnamefont {L.}~\bibnamefont {Xiang}}, \bibinfo {author}
  {\bibfnamefont {T.}~\bibnamefont {Taniguchi}}, \emph {et~al.},\ }\bibfield
  {title} {\bibinfo {title} {Excitonic insulator in a heterojunction moir{\'e}
  superlattice},\ }\href@noop {} {\bibfield  {journal} {\bibinfo  {journal}
  {Nature Physics}\ }\textbf {\bibinfo {volume} {18}},\ \bibinfo {pages} {1171}
  (\bibinfo {year} {2022})}\BibitemShut {NoStop}%
\bibitem [{\citenamefont {Zhang}\ \emph {et~al.}(2022)\citenamefont {Zhang},
  \citenamefont {Regan}, \citenamefont {Wang}, \citenamefont {Zhao},
  \citenamefont {Wang}, \citenamefont {Sayyad}, \citenamefont {Yumigeta},
  \citenamefont {Watanabe}, \citenamefont {Taniguchi}, \citenamefont {Tongay}
  \emph {et~al.}}]{zhang2022correlated}%
  \BibitemOpen
  \bibfield  {author} {\bibinfo {author} {\bibfnamefont {Z.}~\bibnamefont
  {Zhang}}, \bibinfo {author} {\bibfnamefont {E.~C.}\ \bibnamefont {Regan}},
  \bibinfo {author} {\bibfnamefont {D.}~\bibnamefont {Wang}}, \bibinfo {author}
  {\bibfnamefont {W.}~\bibnamefont {Zhao}}, \bibinfo {author} {\bibfnamefont
  {S.}~\bibnamefont {Wang}}, \bibinfo {author} {\bibfnamefont {M.}~\bibnamefont
  {Sayyad}}, \bibinfo {author} {\bibfnamefont {K.}~\bibnamefont {Yumigeta}},
  \bibinfo {author} {\bibfnamefont {K.}~\bibnamefont {Watanabe}}, \bibinfo
  {author} {\bibfnamefont {T.}~\bibnamefont {Taniguchi}}, \bibinfo {author}
  {\bibfnamefont {S.}~\bibnamefont {Tongay}}, \emph {et~al.},\ }\bibfield
  {title} {\bibinfo {title} {{Correlated interlayer exciton insulator in
  heterostructures of monolayer WSe$_2$ and moir{\'e} WS$_2$/WSe$_2$}},\
  }\href@noop {} {\bibfield  {journal} {\bibinfo  {journal} {Nature Physics}\
  }\textbf {\bibinfo {volume} {18}},\ \bibinfo {pages} {1214} (\bibinfo {year}
  {2022})}\BibitemShut {NoStop}%
\bibitem [{\citenamefont {Yankowitz}\ \emph {et~al.}(2019)\citenamefont
  {Yankowitz}, \citenamefont {Ma}, \citenamefont {Jarillo-Herrero},\ and\
  \citenamefont {LeRoy}}]{yankowitz2019van}%
  \BibitemOpen
  \bibfield  {author} {\bibinfo {author} {\bibfnamefont {M.}~\bibnamefont
  {Yankowitz}}, \bibinfo {author} {\bibfnamefont {Q.}~\bibnamefont {Ma}},
  \bibinfo {author} {\bibfnamefont {P.}~\bibnamefont {Jarillo-Herrero}},\ and\
  \bibinfo {author} {\bibfnamefont {B.~J.}\ \bibnamefont {LeRoy}},\ }\bibfield
  {title} {\bibinfo {title} {{van der Waals heterostructures combining graphene
  and hexagonal boron nitride}},\ }\href@noop {} {\bibfield  {journal}
  {\bibinfo  {journal} {Nature Reviews Physics}\ }\textbf {\bibinfo {volume}
  {1}},\ \bibinfo {pages} {112} (\bibinfo {year} {2019})}\BibitemShut {NoStop}%
\bibitem [{\citenamefont {Zhu}\ \emph {et~al.}(2022)\citenamefont {Zhu},
  \citenamefont {Carr}, \citenamefont {Ma},\ and\ \citenamefont
  {Kaxiras}}]{zhu2022electric}%
  \BibitemOpen
  \bibfield  {author} {\bibinfo {author} {\bibfnamefont {Z.}~\bibnamefont
  {Zhu}}, \bibinfo {author} {\bibfnamefont {S.}~\bibnamefont {Carr}}, \bibinfo
  {author} {\bibfnamefont {Q.}~\bibnamefont {Ma}},\ and\ \bibinfo {author}
  {\bibfnamefont {E.}~\bibnamefont {Kaxiras}},\ }\bibfield  {title} {\bibinfo
  {title} {Electric field tunable layer polarization in graphene/boron-nitride
  twisted quadrilayer superlattices},\ }\href@noop {} {\bibfield  {journal}
  {\bibinfo  {journal} {Physical Review B}\ }\textbf {\bibinfo {volume}
  {106}},\ \bibinfo {pages} {205134} (\bibinfo {year} {2022})}\BibitemShut
  {NoStop}%
\bibitem [{\citenamefont {Rozen}\ \emph {et~al.}(2021)\citenamefont {Rozen},
  \citenamefont {Park}, \citenamefont {Zondiner}, \citenamefont {Cao},
  \citenamefont {Rodan-Legrain}, \citenamefont {Taniguchi}, \citenamefont
  {Watanabe}, \citenamefont {Oreg}, \citenamefont {Stern}, \citenamefont {Berg}
  \emph {et~al.}}]{rozen2021entropic}%
  \BibitemOpen
  \bibfield  {author} {\bibinfo {author} {\bibfnamefont {A.}~\bibnamefont
  {Rozen}}, \bibinfo {author} {\bibfnamefont {J.~M.}\ \bibnamefont {Park}},
  \bibinfo {author} {\bibfnamefont {U.}~\bibnamefont {Zondiner}}, \bibinfo
  {author} {\bibfnamefont {Y.}~\bibnamefont {Cao}}, \bibinfo {author}
  {\bibfnamefont {D.}~\bibnamefont {Rodan-Legrain}}, \bibinfo {author}
  {\bibfnamefont {T.}~\bibnamefont {Taniguchi}}, \bibinfo {author}
  {\bibfnamefont {K.}~\bibnamefont {Watanabe}}, \bibinfo {author}
  {\bibfnamefont {Y.}~\bibnamefont {Oreg}}, \bibinfo {author} {\bibfnamefont
  {A.}~\bibnamefont {Stern}}, \bibinfo {author} {\bibfnamefont
  {E.}~\bibnamefont {Berg}}, \emph {et~al.},\ }\bibfield  {title} {\bibinfo
  {title} {Entropic evidence for a pomeranchuk effect in magic-angle
  graphene},\ }\href@noop {} {\bibfield  {journal} {\bibinfo  {journal}
  {Nature}\ }\textbf {\bibinfo {volume} {592}},\ \bibinfo {pages} {214}
  (\bibinfo {year} {2021})}\BibitemShut {NoStop}%
\bibitem [{\citenamefont {Saito}\ \emph {et~al.}(2021)\citenamefont {Saito},
  \citenamefont {Yang}, \citenamefont {Ge}, \citenamefont {Liu}, \citenamefont
  {Taniguchi}, \citenamefont {Watanabe}, \citenamefont {Li}, \citenamefont
  {Berg},\ and\ \citenamefont {Young}}]{saito2021isospin}%
  \BibitemOpen
  \bibfield  {author} {\bibinfo {author} {\bibfnamefont {Y.}~\bibnamefont
  {Saito}}, \bibinfo {author} {\bibfnamefont {F.}~\bibnamefont {Yang}},
  \bibinfo {author} {\bibfnamefont {J.}~\bibnamefont {Ge}}, \bibinfo {author}
  {\bibfnamefont {X.}~\bibnamefont {Liu}}, \bibinfo {author} {\bibfnamefont
  {T.}~\bibnamefont {Taniguchi}}, \bibinfo {author} {\bibfnamefont
  {K.}~\bibnamefont {Watanabe}}, \bibinfo {author} {\bibfnamefont
  {J.}~\bibnamefont {Li}}, \bibinfo {author} {\bibfnamefont {E.}~\bibnamefont
  {Berg}},\ and\ \bibinfo {author} {\bibfnamefont {A.~F.}\ \bibnamefont
  {Young}},\ }\bibfield  {title} {\bibinfo {title} {{Isospin Pomeranchuk effect
  in twisted bilayer graphene}},\ }\href@noop {} {\bibfield  {journal}
  {\bibinfo  {journal} {Nature}\ }\textbf {\bibinfo {volume} {592}},\ \bibinfo
  {pages} {220} (\bibinfo {year} {2021})}\BibitemShut {NoStop}%
\bibitem [{\citenamefont {Park}\ \emph
  {et~al.}(2021{\natexlab{a}})\citenamefont {Park}, \citenamefont {Cao},
  \citenamefont {Watanabe}, \citenamefont {Taniguchi},\ and\ \citenamefont
  {Jarillo-Herrero}}]{park2021tunable}%
  \BibitemOpen
  \bibfield  {author} {\bibinfo {author} {\bibfnamefont {J.~M.}\ \bibnamefont
  {Park}}, \bibinfo {author} {\bibfnamefont {Y.}~\bibnamefont {Cao}}, \bibinfo
  {author} {\bibfnamefont {K.}~\bibnamefont {Watanabe}}, \bibinfo {author}
  {\bibfnamefont {T.}~\bibnamefont {Taniguchi}},\ and\ \bibinfo {author}
  {\bibfnamefont {P.}~\bibnamefont {Jarillo-Herrero}},\ }\bibfield  {title}
  {\bibinfo {title} {Tunable strongly coupled superconductivity in magic-angle
  twisted trilayer graphene},\ }\href@noop {} {\bibfield  {journal} {\bibinfo
  {journal} {Nature}\ }\textbf {\bibinfo {volume} {590}},\ \bibinfo {pages}
  {249} (\bibinfo {year} {2021}{\natexlab{a}})}\BibitemShut {NoStop}%
\bibitem [{\citenamefont {Hao}\ \emph {et~al.}(2021)\citenamefont {Hao},
  \citenamefont {Zimmerman}, \citenamefont {Ledwith}, \citenamefont {Khalaf},
  \citenamefont {Najafabadi}, \citenamefont {Watanabe}, \citenamefont
  {Taniguchi}, \citenamefont {Vishwanath},\ and\ \citenamefont
  {Kim}}]{hao2021electric}%
  \BibitemOpen
  \bibfield  {author} {\bibinfo {author} {\bibfnamefont {Z.}~\bibnamefont
  {Hao}}, \bibinfo {author} {\bibfnamefont {A.}~\bibnamefont {Zimmerman}},
  \bibinfo {author} {\bibfnamefont {P.}~\bibnamefont {Ledwith}}, \bibinfo
  {author} {\bibfnamefont {E.}~\bibnamefont {Khalaf}}, \bibinfo {author}
  {\bibfnamefont {D.~H.}\ \bibnamefont {Najafabadi}}, \bibinfo {author}
  {\bibfnamefont {K.}~\bibnamefont {Watanabe}}, \bibinfo {author}
  {\bibfnamefont {T.}~\bibnamefont {Taniguchi}}, \bibinfo {author}
  {\bibfnamefont {A.}~\bibnamefont {Vishwanath}},\ and\ \bibinfo {author}
  {\bibfnamefont {P.}~\bibnamefont {Kim}},\ }\bibfield  {title} {\bibinfo
  {title} {Electric field--tunable superconductivity in alternating-twist
  magic-angle trilayer graphene},\ }\href@noop {} {\bibfield  {journal}
  {\bibinfo  {journal} {Science}\ }\textbf {\bibinfo {volume} {371}},\ \bibinfo
  {pages} {1133} (\bibinfo {year} {2021})}\BibitemShut {NoStop}%
\bibitem [{\citenamefont {Zhao}\ \emph {et~al.}(2022)\citenamefont {Zhao},
  \citenamefont {Shen}, \citenamefont {Tao}, \citenamefont {Han}, \citenamefont
  {Kang}, \citenamefont {Watanabe}, \citenamefont {Taniguchi}, \citenamefont
  {Mak},\ and\ \citenamefont {Shan}}]{zhao2022gate}%
  \BibitemOpen
  \bibfield  {author} {\bibinfo {author} {\bibfnamefont {W.}~\bibnamefont
  {Zhao}}, \bibinfo {author} {\bibfnamefont {B.}~\bibnamefont {Shen}}, \bibinfo
  {author} {\bibfnamefont {Z.}~\bibnamefont {Tao}}, \bibinfo {author}
  {\bibfnamefont {Z.}~\bibnamefont {Han}}, \bibinfo {author} {\bibfnamefont
  {K.}~\bibnamefont {Kang}}, \bibinfo {author} {\bibfnamefont {K.}~\bibnamefont
  {Watanabe}}, \bibinfo {author} {\bibfnamefont {T.}~\bibnamefont {Taniguchi}},
  \bibinfo {author} {\bibfnamefont {K.~F.}\ \bibnamefont {Mak}},\ and\ \bibinfo
  {author} {\bibfnamefont {J.}~\bibnamefont {Shan}},\ }\bibfield  {title}
  {\bibinfo {title} {{Gate-tunable heavy fermions in a moir$\'e$ Kondo
  lattice}},\ }\href@noop {} {\bibfield  {journal} {\bibinfo  {journal}
  {arXiv:2211.00263}\ } (\bibinfo {year} {2022})}\BibitemShut {NoStop}%
\bibitem [{\citenamefont {Ramires}\ and\ \citenamefont
  {Lado}(2021)}]{ramires2021emulating}%
  \BibitemOpen
  \bibfield  {author} {\bibinfo {author} {\bibfnamefont {A.}~\bibnamefont
  {Ramires}}\ and\ \bibinfo {author} {\bibfnamefont {J.~L.}\ \bibnamefont
  {Lado}},\ }\bibfield  {title} {\bibinfo {title} {Emulating heavy fermions in
  twisted trilayer graphene},\ }\href@noop {} {\bibfield  {journal} {\bibinfo
  {journal} {Physical Review Letters}\ }\textbf {\bibinfo {volume} {127}},\
  \bibinfo {pages} {026401} (\bibinfo {year} {2021})}\BibitemShut {NoStop}%
\bibitem [{\citenamefont {Song}\ and\ \citenamefont
  {Bernevig}(2022)}]{song2022magic}%
  \BibitemOpen
  \bibfield  {author} {\bibinfo {author} {\bibfnamefont {Z.-D.}\ \bibnamefont
  {Song}}\ and\ \bibinfo {author} {\bibfnamefont {B.~A.}\ \bibnamefont
  {Bernevig}},\ }\bibfield  {title} {\bibinfo {title} {Magic-angle twisted
  bilayer graphene as a topological heavy fermion problem},\ }\href@noop {}
  {\bibfield  {journal} {\bibinfo  {journal} {Physical Review Letters}\
  }\textbf {\bibinfo {volume} {129}},\ \bibinfo {pages} {047601} (\bibinfo
  {year} {2022})}\BibitemShut {NoStop}%
\bibitem [{\citenamefont {Kumar}\ \emph {et~al.}(2022)\citenamefont {Kumar},
  \citenamefont {Hu}, \citenamefont {MacDonald},\ and\ \citenamefont
  {Potter}}]{kumar2022gate}%
  \BibitemOpen
  \bibfield  {author} {\bibinfo {author} {\bibfnamefont {A.}~\bibnamefont
  {Kumar}}, \bibinfo {author} {\bibfnamefont {N.~C.}\ \bibnamefont {Hu}},
  \bibinfo {author} {\bibfnamefont {A.~H.}\ \bibnamefont {MacDonald}},\ and\
  \bibinfo {author} {\bibfnamefont {A.~C.}\ \bibnamefont {Potter}},\ }\bibfield
   {title} {\bibinfo {title} {Gate-tunable heavy fermion quantum criticality in
  a moir{\'e} kondo lattice},\ }\href@noop {} {\bibfield  {journal} {\bibinfo
  {journal} {Physical Review B}\ }\textbf {\bibinfo {volume} {106}},\ \bibinfo
  {pages} {L041116} (\bibinfo {year} {2022})}\BibitemShut {NoStop}%
\bibitem [{\citenamefont {Zeng}\ \emph {et~al.}(2022)\citenamefont {Zeng},
  \citenamefont {Wei},\ and\ \citenamefont {MacDonald}}]{zeng2022layer}%
  \BibitemOpen
  \bibfield  {author} {\bibinfo {author} {\bibfnamefont {Y.}~\bibnamefont
  {Zeng}}, \bibinfo {author} {\bibfnamefont {N.}~\bibnamefont {Wei}},\ and\
  \bibinfo {author} {\bibfnamefont {A.~H.}\ \bibnamefont {MacDonald}},\
  }\bibfield  {title} {\bibinfo {title} {Layer pseudospin magnetism in a
  transition metal dichalcogenide double-moir{\'e} system},\ }\href@noop {}
  {\bibfield  {journal} {\bibinfo  {journal} {Physical Review B}\ }\textbf
  {\bibinfo {volume} {106}},\ \bibinfo {pages} {165105} (\bibinfo {year}
  {2022})}\BibitemShut {NoStop}%
\bibitem [{\citenamefont {Taychatanapat}\ and\ \citenamefont
  {Jarillo-Herrero}(2010)}]{taychatanapat2010electronic}%
  \BibitemOpen
  \bibfield  {author} {\bibinfo {author} {\bibfnamefont {T.}~\bibnamefont
  {Taychatanapat}}\ and\ \bibinfo {author} {\bibfnamefont {P.}~\bibnamefont
  {Jarillo-Herrero}},\ }\bibfield  {title} {\bibinfo {title} {Electronic
  transport in dual-gated bilayer graphene at large displacement fields},\
  }\href@noop {} {\bibfield  {journal} {\bibinfo  {journal} {Physical review
  letters}\ }\textbf {\bibinfo {volume} {105}},\ \bibinfo {pages} {166601}
  (\bibinfo {year} {2010})}\BibitemShut {NoStop}%
\bibitem [{\citenamefont {Zheng}\ \emph {et~al.}(2020)\citenamefont {Zheng},
  \citenamefont {Ma}, \citenamefont {Bi}, \citenamefont {de~La~Barrera},
  \citenamefont {Liu}, \citenamefont {Mao}, \citenamefont {Zhang},
  \citenamefont {Kiper}, \citenamefont {Watanabe}, \citenamefont {Taniguchi}
  \emph {et~al.}}]{zheng2020unconventional}%
  \BibitemOpen
  \bibfield  {author} {\bibinfo {author} {\bibfnamefont {Z.}~\bibnamefont
  {Zheng}}, \bibinfo {author} {\bibfnamefont {Q.}~\bibnamefont {Ma}}, \bibinfo
  {author} {\bibfnamefont {Z.}~\bibnamefont {Bi}}, \bibinfo {author}
  {\bibfnamefont {S.}~\bibnamefont {de~La~Barrera}}, \bibinfo {author}
  {\bibfnamefont {M.-H.}\ \bibnamefont {Liu}}, \bibinfo {author} {\bibfnamefont
  {N.}~\bibnamefont {Mao}}, \bibinfo {author} {\bibfnamefont {Y.}~\bibnamefont
  {Zhang}}, \bibinfo {author} {\bibfnamefont {N.}~\bibnamefont {Kiper}},
  \bibinfo {author} {\bibfnamefont {K.}~\bibnamefont {Watanabe}}, \bibinfo
  {author} {\bibfnamefont {T.}~\bibnamefont {Taniguchi}}, \emph {et~al.},\
  }\bibfield  {title} {\bibinfo {title} {Unconventional ferroelectricity in
  moir{\'e} heterostructures},\ }\href@noop {} {\bibfield  {journal} {\bibinfo
  {journal} {Nature}\ }\textbf {\bibinfo {volume} {588}},\ \bibinfo {pages}
  {71} (\bibinfo {year} {2020})}\BibitemShut {NoStop}%
\bibitem [{\citenamefont {Klein}\ \emph {et~al.}(2023)\citenamefont {Klein},
  \citenamefont {Xia}, \citenamefont {MacNeill}, \citenamefont {Watanabe},
  \citenamefont {Taniguchi},\ and\ \citenamefont
  {Jarillo-Herrero}}]{klein2023electrical}%
  \BibitemOpen
  \bibfield  {author} {\bibinfo {author} {\bibfnamefont {D.~R.}\ \bibnamefont
  {Klein}}, \bibinfo {author} {\bibfnamefont {L.-Q.}\ \bibnamefont {Xia}},
  \bibinfo {author} {\bibfnamefont {D.}~\bibnamefont {MacNeill}}, \bibinfo
  {author} {\bibfnamefont {K.}~\bibnamefont {Watanabe}}, \bibinfo {author}
  {\bibfnamefont {T.}~\bibnamefont {Taniguchi}},\ and\ \bibinfo {author}
  {\bibfnamefont {P.}~\bibnamefont {Jarillo-Herrero}},\ }\bibfield  {title}
  {\bibinfo {title} {Electrical switching of a bistable moir{\'e}
  superconductor},\ }\href@noop {} {\bibfield  {journal} {\bibinfo  {journal}
  {Nature Nanotechnology}\ ,\ \bibinfo {pages} {1}} (\bibinfo {year}
  {2023})}\BibitemShut {NoStop}%
\bibitem [{\citenamefont {Niu}\ \emph {et~al.}(2022)\citenamefont {Niu},
  \citenamefont {Li}, \citenamefont {Han}, \citenamefont {Qu}, \citenamefont
  {Ding}, \citenamefont {Wang}, \citenamefont {Liu}, \citenamefont {Liu},
  \citenamefont {Han}, \citenamefont {Watanabe} \emph {et~al.}}]{niu2022giant}%
  \BibitemOpen
  \bibfield  {author} {\bibinfo {author} {\bibfnamefont {R.}~\bibnamefont
  {Niu}}, \bibinfo {author} {\bibfnamefont {Z.}~\bibnamefont {Li}}, \bibinfo
  {author} {\bibfnamefont {X.}~\bibnamefont {Han}}, \bibinfo {author}
  {\bibfnamefont {Z.}~\bibnamefont {Qu}}, \bibinfo {author} {\bibfnamefont
  {D.}~\bibnamefont {Ding}}, \bibinfo {author} {\bibfnamefont {Z.}~\bibnamefont
  {Wang}}, \bibinfo {author} {\bibfnamefont {Q.}~\bibnamefont {Liu}}, \bibinfo
  {author} {\bibfnamefont {T.}~\bibnamefont {Liu}}, \bibinfo {author}
  {\bibfnamefont {C.}~\bibnamefont {Han}}, \bibinfo {author} {\bibfnamefont
  {K.}~\bibnamefont {Watanabe}}, \emph {et~al.},\ }\bibfield  {title} {\bibinfo
  {title} {Giant ferroelectric polarization in a bilayer graphene
  heterostructure},\ }\href@noop {} {\bibfield  {journal} {\bibinfo  {journal}
  {Nature Communications}\ }\textbf {\bibinfo {volume} {13}},\ \bibinfo {pages}
  {1} (\bibinfo {year} {2022})}\BibitemShut {NoStop}%
\bibitem [{\citenamefont {Zondiner}\ \emph {et~al.}(2020)\citenamefont
  {Zondiner}, \citenamefont {Rozen}, \citenamefont {Rodan-Legrain},
  \citenamefont {Cao}, \citenamefont {Queiroz}, \citenamefont {Taniguchi},
  \citenamefont {Watanabe}, \citenamefont {Oreg}, \citenamefont {von Oppen},
  \citenamefont {Stern} \emph {et~al.}}]{zondiner2020cascade}%
  \BibitemOpen
  \bibfield  {author} {\bibinfo {author} {\bibfnamefont {U.}~\bibnamefont
  {Zondiner}}, \bibinfo {author} {\bibfnamefont {A.}~\bibnamefont {Rozen}},
  \bibinfo {author} {\bibfnamefont {D.}~\bibnamefont {Rodan-Legrain}}, \bibinfo
  {author} {\bibfnamefont {Y.}~\bibnamefont {Cao}}, \bibinfo {author}
  {\bibfnamefont {R.}~\bibnamefont {Queiroz}}, \bibinfo {author} {\bibfnamefont
  {T.}~\bibnamefont {Taniguchi}}, \bibinfo {author} {\bibfnamefont
  {K.}~\bibnamefont {Watanabe}}, \bibinfo {author} {\bibfnamefont
  {Y.}~\bibnamefont {Oreg}}, \bibinfo {author} {\bibfnamefont {F.}~\bibnamefont
  {von Oppen}}, \bibinfo {author} {\bibfnamefont {A.}~\bibnamefont {Stern}},
  \emph {et~al.},\ }\bibfield  {title} {\bibinfo {title} {Cascade of phase
  transitions and dirac revivals in magic-angle graphene},\ }\href@noop {}
  {\bibfield  {journal} {\bibinfo  {journal} {Nature}\ }\textbf {\bibinfo
  {volume} {582}},\ \bibinfo {pages} {203} (\bibinfo {year}
  {2020})}\BibitemShut {NoStop}%
\bibitem [{\citenamefont {Wong}\ \emph {et~al.}(2020)\citenamefont {Wong},
  \citenamefont {Nuckolls}, \citenamefont {Oh}, \citenamefont {Lian},
  \citenamefont {Xie}, \citenamefont {Jeon}, \citenamefont {Watanabe},
  \citenamefont {Taniguchi}, \citenamefont {Bernevig},\ and\ \citenamefont
  {Yazdani}}]{wong2020cascade}%
  \BibitemOpen
  \bibfield  {author} {\bibinfo {author} {\bibfnamefont {D.}~\bibnamefont
  {Wong}}, \bibinfo {author} {\bibfnamefont {K.~P.}\ \bibnamefont {Nuckolls}},
  \bibinfo {author} {\bibfnamefont {M.}~\bibnamefont {Oh}}, \bibinfo {author}
  {\bibfnamefont {B.}~\bibnamefont {Lian}}, \bibinfo {author} {\bibfnamefont
  {Y.}~\bibnamefont {Xie}}, \bibinfo {author} {\bibfnamefont {S.}~\bibnamefont
  {Jeon}}, \bibinfo {author} {\bibfnamefont {K.}~\bibnamefont {Watanabe}},
  \bibinfo {author} {\bibfnamefont {T.}~\bibnamefont {Taniguchi}}, \bibinfo
  {author} {\bibfnamefont {B.~A.}\ \bibnamefont {Bernevig}},\ and\ \bibinfo
  {author} {\bibfnamefont {A.}~\bibnamefont {Yazdani}},\ }\bibfield  {title}
  {\bibinfo {title} {Cascade of electronic transitions in magic-angle twisted
  bilayer graphene},\ }\href@noop {} {\bibfield  {journal} {\bibinfo  {journal}
  {Nature}\ }\textbf {\bibinfo {volume} {582}},\ \bibinfo {pages} {198}
  (\bibinfo {year} {2020})}\BibitemShut {NoStop}%
\bibitem [{\citenamefont {Kim}\ \emph {et~al.}(2012)\citenamefont {Kim},
  \citenamefont {Jo}, \citenamefont {Dillen}, \citenamefont {Ferrer},
  \citenamefont {Fallahazad}, \citenamefont {Yao}, \citenamefont {Banerjee},\
  and\ \citenamefont {Tutuc}}]{kim2012direct}%
  \BibitemOpen
  \bibfield  {author} {\bibinfo {author} {\bibfnamefont {S.}~\bibnamefont
  {Kim}}, \bibinfo {author} {\bibfnamefont {I.}~\bibnamefont {Jo}}, \bibinfo
  {author} {\bibfnamefont {D.}~\bibnamefont {Dillen}}, \bibinfo {author}
  {\bibfnamefont {D.}~\bibnamefont {Ferrer}}, \bibinfo {author} {\bibfnamefont
  {B.}~\bibnamefont {Fallahazad}}, \bibinfo {author} {\bibfnamefont
  {Z.}~\bibnamefont {Yao}}, \bibinfo {author} {\bibfnamefont {S.}~\bibnamefont
  {Banerjee}},\ and\ \bibinfo {author} {\bibfnamefont {E.}~\bibnamefont
  {Tutuc}},\ }\bibfield  {title} {\bibinfo {title} {{Direct measurement of the
  Fermi energy in graphene using a double-layer heterostructure}},\ }\href@noop
  {} {\bibfield  {journal} {\bibinfo  {journal} {Physical review letters}\
  }\textbf {\bibinfo {volume} {108}},\ \bibinfo {pages} {116404} (\bibinfo
  {year} {2012})}\BibitemShut {NoStop}%
\bibitem [{\citenamefont {Park}\ \emph
  {et~al.}(2021{\natexlab{b}})\citenamefont {Park}, \citenamefont {Cao},
  \citenamefont {Watanabe}, \citenamefont {Taniguchi},\ and\ \citenamefont
  {Jarillo-Herrero}}]{park2021flavour}%
  \BibitemOpen
  \bibfield  {author} {\bibinfo {author} {\bibfnamefont {J.~M.}\ \bibnamefont
  {Park}}, \bibinfo {author} {\bibfnamefont {Y.}~\bibnamefont {Cao}}, \bibinfo
  {author} {\bibfnamefont {K.}~\bibnamefont {Watanabe}}, \bibinfo {author}
  {\bibfnamefont {T.}~\bibnamefont {Taniguchi}},\ and\ \bibinfo {author}
  {\bibfnamefont {P.}~\bibnamefont {Jarillo-Herrero}},\ }\bibfield  {title}
  {\bibinfo {title} {Flavour hund’s coupling, chern gaps and charge
  diffusivity in moir{\'e} graphene},\ }\href@noop {} {\bibfield  {journal}
  {\bibinfo  {journal} {Nature}\ }\textbf {\bibinfo {volume} {592}},\ \bibinfo
  {pages} {43} (\bibinfo {year} {2021}{\natexlab{b}})}\BibitemShut {NoStop}%
\bibitem [{\citenamefont {Fei}\ \emph {et~al.}(2018)\citenamefont {Fei},
  \citenamefont {Zhao}, \citenamefont {Palomaki}, \citenamefont {Sun},
  \citenamefont {Miller}, \citenamefont {Zhao}, \citenamefont {Yan},
  \citenamefont {Xu},\ and\ \citenamefont {Cobden}}]{fei2018ferroelectric}%
  \BibitemOpen
  \bibfield  {author} {\bibinfo {author} {\bibfnamefont {Z.}~\bibnamefont
  {Fei}}, \bibinfo {author} {\bibfnamefont {W.}~\bibnamefont {Zhao}}, \bibinfo
  {author} {\bibfnamefont {T.~A.}\ \bibnamefont {Palomaki}}, \bibinfo {author}
  {\bibfnamefont {B.}~\bibnamefont {Sun}}, \bibinfo {author} {\bibfnamefont
  {M.~K.}\ \bibnamefont {Miller}}, \bibinfo {author} {\bibfnamefont
  {Z.}~\bibnamefont {Zhao}}, \bibinfo {author} {\bibfnamefont {J.}~\bibnamefont
  {Yan}}, \bibinfo {author} {\bibfnamefont {X.}~\bibnamefont {Xu}},\ and\
  \bibinfo {author} {\bibfnamefont {D.~H.}\ \bibnamefont {Cobden}},\ }\bibfield
   {title} {\bibinfo {title} {Ferroelectric switching of a two-dimensional
  metal},\ }\href@noop {} {\bibfield  {journal} {\bibinfo  {journal} {Nature}\
  }\textbf {\bibinfo {volume} {560}},\ \bibinfo {pages} {336} (\bibinfo {year}
  {2018})}\BibitemShut {NoStop}%
\bibitem [{\citenamefont {Sharma}\ \emph {et~al.}(2019)\citenamefont {Sharma},
  \citenamefont {Xiang}, \citenamefont {Shao}, \citenamefont {Zhang},
  \citenamefont {Tsymbal}, \citenamefont {Hamilton},\ and\ \citenamefont
  {Seidel}}]{sharma2019room}%
  \BibitemOpen
  \bibfield  {author} {\bibinfo {author} {\bibfnamefont {P.}~\bibnamefont
  {Sharma}}, \bibinfo {author} {\bibfnamefont {F.-X.}\ \bibnamefont {Xiang}},
  \bibinfo {author} {\bibfnamefont {D.-F.}\ \bibnamefont {Shao}}, \bibinfo
  {author} {\bibfnamefont {D.}~\bibnamefont {Zhang}}, \bibinfo {author}
  {\bibfnamefont {E.~Y.}\ \bibnamefont {Tsymbal}}, \bibinfo {author}
  {\bibfnamefont {A.~R.}\ \bibnamefont {Hamilton}},\ and\ \bibinfo {author}
  {\bibfnamefont {J.}~\bibnamefont {Seidel}},\ }\bibfield  {title} {\bibinfo
  {title} {A room-temperature ferroelectric semimetal},\ }\href@noop {}
  {\bibfield  {journal} {\bibinfo  {journal} {Science advances}\ }\textbf
  {\bibinfo {volume} {5}},\ \bibinfo {pages} {eaax5080} (\bibinfo {year}
  {2019})}\BibitemShut {NoStop}%
\bibitem [{\citenamefont {de~la Barrera}\ \emph {et~al.}(2021)\citenamefont
  {de~la Barrera}, \citenamefont {Cao}, \citenamefont {Gao}, \citenamefont
  {Gao}, \citenamefont {Bheemarasetty}, \citenamefont {Yan}, \citenamefont
  {Mandrus}, \citenamefont {Zhu}, \citenamefont {Xiao},\ and\ \citenamefont
  {Hunt}}]{de2021direct}%
  \BibitemOpen
  \bibfield  {author} {\bibinfo {author} {\bibfnamefont {S.~C.}\ \bibnamefont
  {de~la Barrera}}, \bibinfo {author} {\bibfnamefont {Q.}~\bibnamefont {Cao}},
  \bibinfo {author} {\bibfnamefont {Y.}~\bibnamefont {Gao}}, \bibinfo {author}
  {\bibfnamefont {Y.}~\bibnamefont {Gao}}, \bibinfo {author} {\bibfnamefont
  {V.~S.}\ \bibnamefont {Bheemarasetty}}, \bibinfo {author} {\bibfnamefont
  {J.}~\bibnamefont {Yan}}, \bibinfo {author} {\bibfnamefont {D.~G.}\
  \bibnamefont {Mandrus}}, \bibinfo {author} {\bibfnamefont {W.}~\bibnamefont
  {Zhu}}, \bibinfo {author} {\bibfnamefont {D.}~\bibnamefont {Xiao}},\ and\
  \bibinfo {author} {\bibfnamefont {B.~M.}\ \bibnamefont {Hunt}},\ }\bibfield
  {title} {\bibinfo {title} {Direct measurement of ferroelectric polarization
  in a tunable semimetal},\ }\href@noop {} {\bibfield  {journal} {\bibinfo
  {journal} {Nature Communications}\ }\textbf {\bibinfo {volume} {12}},\
  \bibinfo {pages} {5298} (\bibinfo {year} {2021})}\BibitemShut {NoStop}%
\bibitem [{\citenamefont {Yang}\ and\ \citenamefont
  {Wu}(2023)}]{yang2023across}%
  \BibitemOpen
  \bibfield  {author} {\bibinfo {author} {\bibfnamefont {L.}~\bibnamefont
  {Yang}}\ and\ \bibinfo {author} {\bibfnamefont {M.}~\bibnamefont {Wu}},\
  }\bibfield  {title} {\bibinfo {title} {Across-layer sliding ferroelectricity
  in 2d heterolayers},\ }\href@noop {} {\bibfield  {journal} {\bibinfo
  {journal} {Advanced Functional Materials}\ ,\ \bibinfo {pages} {2301105}}
  (\bibinfo {year} {2023})}\BibitemShut {NoStop}%
\bibitem [{\citenamefont {Yasuda}\ \emph {et~al.}(2021)\citenamefont {Yasuda},
  \citenamefont {Wang}, \citenamefont {Watanabe}, \citenamefont {Taniguchi},\
  and\ \citenamefont {Jarillo-Herrero}}]{Yasuda2021}%
  \BibitemOpen
  \bibfield  {author} {\bibinfo {author} {\bibfnamefont {K.}~\bibnamefont
  {Yasuda}}, \bibinfo {author} {\bibfnamefont {X.}~\bibnamefont {Wang}},
  \bibinfo {author} {\bibfnamefont {K.}~\bibnamefont {Watanabe}}, \bibinfo
  {author} {\bibfnamefont {T.}~\bibnamefont {Taniguchi}},\ and\ \bibinfo
  {author} {\bibfnamefont {P.}~\bibnamefont {Jarillo-Herrero}},\ }\bibfield
  {title} {\bibinfo {title} {Stacking-engineered ferroelectricity in bilayer
  boron nitride},\ }\href@noop {} {\bibfield  {journal} {\bibinfo  {journal}
  {Science}\ }\textbf {\bibinfo {volume} {372}},\ \bibinfo {pages} {1458}
  (\bibinfo {year} {2021})}\BibitemShut {NoStop}%
\bibitem [{\citenamefont {Gonzalo}\ and\ \citenamefont
  {Jim{\'e}nez}(2005)}]{gonzalo2005ferroelectricity}%
  \BibitemOpen
  \bibfield  {author} {\bibinfo {author} {\bibfnamefont {J.~A.}\ \bibnamefont
  {Gonzalo}}\ and\ \bibinfo {author} {\bibfnamefont {B.}~\bibnamefont
  {Jim{\'e}nez}},\ }\href@noop {} {\emph {\bibinfo {title} {Ferroelectricity:
  the fundamentals collection}}},\ Vol.~\bibinfo {volume} {10}\ (\bibinfo
  {publisher} {John Wiley \& Sons},\ \bibinfo {year} {2005})\BibitemShut
  {NoStop}%
\bibitem [{\citenamefont {Preisach}(1935)}]{preisach1935magnetische}%
  \BibitemOpen
  \bibfield  {author} {\bibinfo {author} {\bibfnamefont {F.}~\bibnamefont
  {Preisach}},\ }\bibfield  {title} {\bibinfo {title} {{\"U}ber die magnetische
  {N}achwirkung},\ }\href@noop {} {\bibfield  {journal} {\bibinfo  {journal}
  {Zeitschrift f{\"u}r physik}\ }\textbf {\bibinfo {volume} {94}},\ \bibinfo
  {pages} {277} (\bibinfo {year} {1935})}\BibitemShut {NoStop}%
\bibitem [{\citenamefont {Cima}\ \emph {et~al.}(2002)\citenamefont {Cima},
  \citenamefont {Laboure},\ and\ \citenamefont
  {Muralt}}]{cima2002characterization}%
  \BibitemOpen
  \bibfield  {author} {\bibinfo {author} {\bibfnamefont {L.}~\bibnamefont
  {Cima}}, \bibinfo {author} {\bibfnamefont {E.}~\bibnamefont {Laboure}},\ and\
  \bibinfo {author} {\bibfnamefont {P.}~\bibnamefont {Muralt}},\ }\bibfield
  {title} {\bibinfo {title} {{Characterization and model of ferroelectrics
  based on experimental Preisach density}},\ }\href@noop {} {\bibfield
  {journal} {\bibinfo  {journal} {Review of scientific instruments}\ }\textbf
  {\bibinfo {volume} {73}},\ \bibinfo {pages} {3546} (\bibinfo {year}
  {2002})}\BibitemShut {NoStop}%
\bibitem [{\citenamefont {Ribeiro-Palau}\ \emph {et~al.}(2018)\citenamefont
  {Ribeiro-Palau}, \citenamefont {Zhang}, \citenamefont {Watanabe},
  \citenamefont {Taniguchi}, \citenamefont {Hone},\ and\ \citenamefont
  {Dean}}]{ribeiro2018twistable}%
  \BibitemOpen
  \bibfield  {author} {\bibinfo {author} {\bibfnamefont {R.}~\bibnamefont
  {Ribeiro-Palau}}, \bibinfo {author} {\bibfnamefont {C.}~\bibnamefont
  {Zhang}}, \bibinfo {author} {\bibfnamefont {K.}~\bibnamefont {Watanabe}},
  \bibinfo {author} {\bibfnamefont {T.}~\bibnamefont {Taniguchi}}, \bibinfo
  {author} {\bibfnamefont {J.}~\bibnamefont {Hone}},\ and\ \bibinfo {author}
  {\bibfnamefont {C.~R.}\ \bibnamefont {Dean}},\ }\bibfield  {title} {\bibinfo
  {title} {Twistable electronics with dynamically rotatable heterostructures},\
  }\href@noop {} {\bibfield  {journal} {\bibinfo  {journal} {Science}\ }\textbf
  {\bibinfo {volume} {361}},\ \bibinfo {pages} {690} (\bibinfo {year}
  {2018})}\BibitemShut {NoStop}%
\bibitem [{\citenamefont {Inbar}\ \emph {et~al.}(2023)\citenamefont {Inbar},
  \citenamefont {Birkbeck}, \citenamefont {Xiao}, \citenamefont {Taniguchi},
  \citenamefont {Watanabe}, \citenamefont {Yan}, \citenamefont {Oreg},
  \citenamefont {Stern}, \citenamefont {Berg},\ and\ \citenamefont
  {Ilani}}]{inbar2023quantum}%
  \BibitemOpen
  \bibfield  {author} {\bibinfo {author} {\bibfnamefont {A.}~\bibnamefont
  {Inbar}}, \bibinfo {author} {\bibfnamefont {J.}~\bibnamefont {Birkbeck}},
  \bibinfo {author} {\bibfnamefont {J.}~\bibnamefont {Xiao}}, \bibinfo {author}
  {\bibfnamefont {T.}~\bibnamefont {Taniguchi}}, \bibinfo {author}
  {\bibfnamefont {K.}~\bibnamefont {Watanabe}}, \bibinfo {author}
  {\bibfnamefont {B.}~\bibnamefont {Yan}}, \bibinfo {author} {\bibfnamefont
  {Y.}~\bibnamefont {Oreg}}, \bibinfo {author} {\bibfnamefont {A.}~\bibnamefont
  {Stern}}, \bibinfo {author} {\bibfnamefont {E.}~\bibnamefont {Berg}},\ and\
  \bibinfo {author} {\bibfnamefont {S.}~\bibnamefont {Ilani}},\ }\bibfield
  {title} {\bibinfo {title} {The quantum twisting microscope},\ }\href@noop {}
  {\bibfield  {journal} {\bibinfo  {journal} {Nature}\ }\textbf {\bibinfo
  {volume} {614}},\ \bibinfo {pages} {682} (\bibinfo {year}
  {2023})}\BibitemShut {NoStop}%
\bibitem [{\citenamefont {Li}\ \emph {et~al.}(2021)\citenamefont {Li},
  \citenamefont {Li}, \citenamefont {Regan}, \citenamefont {Wang},
  \citenamefont {Zhao}, \citenamefont {Kahn}, \citenamefont {Yumigeta},
  \citenamefont {Blei}, \citenamefont {Taniguchi}, \citenamefont {Watanabe}
  \emph {et~al.}}]{li2021imaging}%
  \BibitemOpen
  \bibfield  {author} {\bibinfo {author} {\bibfnamefont {H.}~\bibnamefont
  {Li}}, \bibinfo {author} {\bibfnamefont {S.}~\bibnamefont {Li}}, \bibinfo
  {author} {\bibfnamefont {E.~C.}\ \bibnamefont {Regan}}, \bibinfo {author}
  {\bibfnamefont {D.}~\bibnamefont {Wang}}, \bibinfo {author} {\bibfnamefont
  {W.}~\bibnamefont {Zhao}}, \bibinfo {author} {\bibfnamefont {S.}~\bibnamefont
  {Kahn}}, \bibinfo {author} {\bibfnamefont {K.}~\bibnamefont {Yumigeta}},
  \bibinfo {author} {\bibfnamefont {M.}~\bibnamefont {Blei}}, \bibinfo {author}
  {\bibfnamefont {T.}~\bibnamefont {Taniguchi}}, \bibinfo {author}
  {\bibfnamefont {K.}~\bibnamefont {Watanabe}}, \emph {et~al.},\ }\bibfield
  {title} {\bibinfo {title} {Imaging two-dimensional generalized wigner
  crystals},\ }\href@noop {} {\bibfield  {journal} {\bibinfo  {journal}
  {Nature}\ }\textbf {\bibinfo {volume} {597}},\ \bibinfo {pages} {650}
  (\bibinfo {year} {2021})}\BibitemShut {NoStop}%
\bibitem [{\citenamefont {Yan}\ \emph {et~al.}()\citenamefont {Yan},
  \citenamefont {Zheng}, \citenamefont {Sangwan}, \citenamefont {Qian},
  \citenamefont {Wang}, \citenamefont {Liu}, \citenamefont {Watanabe},
  \citenamefont {Taniguchi}, \citenamefont {Xu}, \citenamefont
  {Jarillo-Herrero}, \citenamefont {Ma},\ and\ \citenamefont
  {Hersam}}]{Yan2023The}%
  \BibitemOpen
  \bibfield  {author} {\bibinfo {author} {\bibfnamefont {X.}~\bibnamefont
  {Yan}}, \bibinfo {author} {\bibfnamefont {Z.}~\bibnamefont {Zheng}}, \bibinfo
  {author} {\bibfnamefont {V.~K.}\ \bibnamefont {Sangwan}}, \bibinfo {author}
  {\bibfnamefont {J.~H.}\ \bibnamefont {Qian}}, \bibinfo {author}
  {\bibfnamefont {X.}~\bibnamefont {Wang}}, \bibinfo {author} {\bibfnamefont
  {S.~E.}\ \bibnamefont {Liu}}, \bibinfo {author} {\bibfnamefont
  {K.}~\bibnamefont {Watanabe}}, \bibinfo {author} {\bibfnamefont
  {T.}~\bibnamefont {Taniguchi}}, \bibinfo {author} {\bibfnamefont {S.-Y.}\
  \bibnamefont {Xu}}, \bibinfo {author} {\bibfnamefont {P.}~\bibnamefont
  {Jarillo-Herrero}}, \bibinfo {author} {\bibfnamefont {Q.}~\bibnamefont
  {Ma}},\ and\ \bibinfo {author} {\bibfnamefont {M.~C.}\ \bibnamefont
  {Hersam}},\ }\bibfield  {title} {\bibinfo {title} {{The Moiré Synaptic
  Transistor: A Room-Temperature Quantum Device with Reconfigurable
  Neuromorphic Functionality}},\ }\href@noop {} {\bibinfo  {journal} {This
  paper is under review. It is available upon request to editor}\ }\BibitemShut
  {NoStop}%
\end{thebibliography}%


%

\clearpage
\subsection*{Methods}
\textbf{Device fabrication and rotational alignment:} The BN-BLG-BN stack was made via the standard dry pickup/transfer technique with a polydimethylsiloxane/poly(bisphenol A carbonate) polymer stamp. There are two ways that we make electric contacts. Our consistent observation across different devices indicates that the contact method does not correlate with our device performance. For the top contact method: we first used electron beam lithography to define electrode areas, then etched the top BN within those areas and evaporated Cr/PdAu on top of the exposed graphene. For the bottom contact method: we first used electron beam lithography to define electrode areas, then etched the bottom BN within those areas and evaporated Cr/PdAu. Then bottom BN and contact were annealed at $300^\circ$C for at least 3hrs with $\mathrm{H_2}$/Ar gas before we transferred the top BN and BLG on top. The number of graphene layers was identified through Raman spectroscopy.

We took several approaches to control and determine the rotational alignment between the top (bottom) BN and graphene for Device D1. First, the top and bottom BN pieces originated from a single BN flake, which was pre-etched into two halves using the reactive ion etching (RIE) prior to the pickup/transfer process. This way, we can precisely control the relative angle between the top and bottom BNs (there will still be ambiguity between $0\degree$ and $180\degree$ due to the unknown number of layers). Second, we chose graphene and BN flakes with long and straight edges, which often correspond to armchair or zigzag edges. We then performed the straight edge alignment: The straight edge of the top BN is optically aligned with the straight edge of the bilayer graphene. The bottom BN is subsequently picked up with a rotation of $15\degree$ relative to the top BN. Third, we confirmed the crystallographic orientation of BN and graphene in Device D1 through the optical second harmonic generation, plotted in extended data Figure~\ref{SHG}. The top and bottom BN were misaligned by 15$\degree$, as designed. We also identified the rotational angle of the BLG from a thicker graphite piece attached to it. The crystalline axis of BLG was confirmed to be closely aligned with that of top BN.

Extended data Fig.~\ref{hyssummary} summarizes the typical transport behavior and angle alignment for several devices. All these devices show consistent cascade behaviors with an alternative pattern between anomalous screening and ratcheting regimes.\\

\textbf{Estimate of chemical potential:} We elaborate on our procedure to extract the chemical potential of the localized electron system through an electrostatic model~\cite{kim2012direct,park2021flavour}. Based on our observation, we have established that the staircase transition of the charge-neutral peak is a result of the screening effect from the layer-specific localized system. We can then calculate the chemical potential and carrier density of the localized subsystem using the following electrostatic relations. At the origin, the two subsystems are both at their charge neutrality. When the system is in equilibrium, the electrochemical potential of the two subsystems must be the same. Therefore, we have the following relations:

\begin{equation}
ev_{\textrm{b}}=eV_{\textrm{BG}}-\mu_{\textrm{I}}(n_{\textrm{I}}),
\label{b}
\end{equation}
\begin{equation}
ev_{\textrm{t}}=eV_{\textrm{TG}}-\mu_{\textrm{L}}(n_{\textrm{L}}),
\label{t}
\end{equation}
\begin{equation}
ev_{\textrm{0}}=\mu_{\textrm{L}}(n_{\textrm{L}})-\mu_{\textrm{I}}(n_{\textrm{I}}),
\label{0}
\end{equation}
\begin{equation}
en_{\textrm{L}}=C_{\textrm{TG}}v_{\textrm{t}}-C_{\textrm{0}}v_{\textrm{0}},
\label{enloc}
\end{equation}
\begin{equation}
en_{\textrm{I}}=C_{\textrm{BG}}v_{\textrm{b}}+C_{\textrm{0}}v_{\textrm{0}},
\label{eniti}
\end{equation}

where $v_{\textrm{t}}, v_{\textrm{b}}, v_{\textrm{0}}$ are the relative electric potential from localized electron system to top gate, from itinerant electron system to bottom gate, and from itinerant electron system to localized electron system respectively. $C_{\textrm{BG}}, C_{\textrm{TG}}, C_{\textrm{0}}$ are the geometric capacitance per unit area of the bottom, top dielectric layers, and middle vacuum gap. Here, we take the z offset between the localized and itinerant electron system to be a vacuum gap of 1 \AA. By combining the five relations and keeping the itinerant electron system at charge neutrality ($n_{\textrm{I}}=0$), we can derive the expression for the density and chemical potential of the localized electron system:
\begin{equation}
n_{\textrm{L}}=\frac{C_{\textrm{TG}}V_{\textrm{TG}}}{e}+\frac{(C_{\textrm{TG}}+C_{\textrm{0}})C_{\textrm{BG}}V_{\textrm{BG}}}{eC_{\textrm{0}}},
\label{nlocal}
\end{equation}
\begin{equation}
\mu_{\textrm{L}}=-\frac{eC_{\textrm{BG}}V_{\textrm{BG}}}{C_{\textrm{0}}},
\label{nlocal}
\end{equation}
Therefore, by tracing out the charge neutrality peak of the itinerant electron system in the dual gate map, we can calculate the density and chemical potential of the localized electron system. We would like to point out that we have extracted $n_\mathrm{L}$ through two independent methods. One is based on the relation $n_\mathrm{L} = n_\mathrm{Total} - n_\mathrm{H}$. The other one is based on electrostatic relations that take into account the vacuum gap between the two subsystems. Both results agree nicely with each other.\\

\noindent \textbf{Single-particle band structures:} The band structure calculation for the graphene/BN heterostructure is performed using a low-energy momentum-space continuum model. The Hamiltonian can be written as a $4\times4$ matrix that consists of the intralayers terms (top BN, two graphene sheets, bottom BN) and the interlayer hopping terms. The different twist angles at the top BN/graphene interface ($\theta_\mathrm{t}$) and the bottom BN/graphene interface ($\theta_\mathrm{b}$) are reflected in the interlayer hopping terms (see SI.B for details). The result of $\theta_\mathrm{t} = 0 \degree$ and $\theta_\mathrm{b} = 15 \degree$ is shown in Extended Data Fig.~\ref{calculation}\textbf{a}. Interestingly, the layer asymmetric moir\'e potential leads to a gap-opening behavior at zero electric fields, which flattens the band dispersion at the low energy and highly polarizes the electronic states at the conduction and valence band edges towards different layers. In our calculation (Extended Data Fig.~\ref{calculation}\textbf{a}), the conduction band bottom corresponds to the top-layer character while the valence band top corresponds to the bottom-layer character. In our calculation, the BN band extrema are far from the low-energy features of interest near the graphene charge neutral point. Therefore, the layer character is primarily associated with the top and bottom graphene sheets. Applying the perpendicular electric field can close and re-open the gap and flip the layer characters associated with the valence and conduction bands (see SI.B).\\

\noindent \textbf{Real-space wavefunction distribution:} Due to the layer-specific moir\'e potential, the layer-polarized electronic states acquire distinct wavefunction distribution in real space. Since the top BN is closely aligned with the BLG and the bottom BN is misaligned by a large angle, we expect only the electronic states polarized to the top graphene sheet experience long-range potential modulation. To confirm this idea, we perform an inverse Fourier transform that sums over the wavefunction weights that correspond to each momentum degree of freedom $\vec{q}^{(\ell)}$ on the $l$ layer ($l$ stands for top or bottom):
\begin{eqnarray}
  \Psi^\ell_{n, \vec{q}} (\vec{r}) = \sum_{\alpha = A,B} \sum_{\vec{q}^{(\ell)}} \psi_{n, \vec{q}, \alpha} (\vec{q}^{(\ell)}) e^{- i \vec{q}^{(\ell)}\cdot \vec{r} },
\end{eqnarray}
where $n$ is the band index, $\vec{q}$ is the center site of the Hamiltonian, and $\psi_\alpha (\vec{q}^{(\ell)}) $ is the wavefunction weight that corresponds to momentum $\vec{q}^{(\ell)}$ and sublattice $\alpha$. From this (see SI.I for details), we obtain the layer-projected real-space wavefunction distribution at a given position $\vec{r}$. In correspondence with the calculated band structure in Extended Data Fig.~\ref{calculation}\textbf{a}, Extended Data Fig.~\ref{calculation}\textbf{b} shows the real-space wavefunction distribution calculated for the conduction and valence band edges. The conductance band bottom, which has a top layer character, exhibits the long-wavelength modulation of the wavefunction distribution and partial electron localization at the AA-like sites. In contrast, the valence band top, which has a bottom layer character, does not exhibit the long-range modulation. This striking difference in real space is surprising as the band dispersion looks similar for the conduction and valence band edges. Our calculation demonstrates that each layer polarized states show electron localization at the scale of their corresponding moir\'e superlattice: For top layer polarized states, the potential modulation reflects the moir\'e periodicity. For bottom layer polarized states, the moir\'e wavelength is too small such that the wavefunction distribution is essentially uniform.\\

The influence of layer-specific moir\'e potential on the real-space wavefunction distribution is further confirmed by calculation on different angle combinations. As shown in Extended Data Fig.~\ref{WF}, we calculated the real-space wavefunction distribution near the conduction and valance band edges for angle combination $\theta_\mathrm{t}=0^{\circ}, \theta_\mathrm{b}=15^{\circ}$;  $\theta_\mathrm{t}=0^{\circ}, \theta_\mathrm{b}=30^{\circ}$; and $\theta_\mathrm{t}=5^{\circ}, \theta_\mathrm{b}=10^{\circ}$. In each angle alignment scenario, the polarized real-space wavefunction distribution always reflects the moir\'e potential length scale of their respective layer, which points to a layer-specific Coulomb interaction where one layer shows a dominating effect. In Extended Data Fig.~\ref{WF_Ddep}, we further show that as the perpendicular electric field flips the layer polarization of the valence and conduction bands, their real-space wavefunction distribution also flips, always matching the moir\'e potential length scale of their respective layer.\\

\noindent \textbf{Construction of localized and itinerant subsystems:} From the real-space wavefunction distribution, we can decompose our BLG/BN moir\'e system into a top layer-polarized localized subsystem and a BLG-like layer-polarizable itinerant subsystem. The layer-polarized localized subsystem corresponds to the confined wavefunction distribution at the top layer, which is about 20$\%$ of the total wavefunction weight at that layer according to Extended Data Fig.~\ref{calculation}\textbf{b}. The BLG-like layer-polarizable itinerant subsystem is the rest uniform background at the top layer combined with the states at the bottom layer. We can see from the single-particle calculation that upon applying the perpendicular electric field, the top layer-polarized localized subsystem can turn from being electron-like to hole-like. The layer polarization of the uniform background can also be reversed. The single-particle electronic structure supports the existence of the two subsystems that can both be electrically tuned.\\

\clearpage

\textbf{Acknowledgements:} The authors would like to acknowledge helpful discussions with Kenneth Burch, Nick Bultinck, Yu He, Steve Kivelson, Leonid Levitov, Allan Macdonald, Ivar Martin, Sid Parameswaran, Ying Ran, Justin Song, Steve Simon, Chandra Varma, and Yonglong Xie. We also thank Houchen Li for his assistance in the optical SHG measurement. This work was primarily supported by the Center for the Advancement of Topological Semimetals, an Energy Frontier Research Center funded by the US Department of Energy Office of Science, through the Ames Laboratory under contract DE-AC02-07CH11358 (data analysis and manuscript writing), by the Gordon and Betty Moore Foundations EPiQS Initiative through Grant GBMF9643 to PJH, and by the Ramon Areces Foundation. PJH and QM acknowledge partial support by the AFOSR grant FA9550-21-1-0319 (device fabrication and transport measurements). This work was performed in part at the Harvard University Center for Nanoscale Systems (CNS), a member of the National Nanotechnology Coordinated Infrastructure Network (NNCI), which is supported by the National Science Foundation under NSF ECCS award no.1541959. QM and LF also acknowledges support from the NSF Convergence program (NSF ITE-2235945) and the CIFAR program. ZRZ acknowledges support from the National Science Foundation Graduate Research Fellowship under Grant No. 2141064. EK acknowledges support from a Simons Foundation Award no. 896626. KW and TT acknowledge support from JSPS KAKENHI (Grant Numbers 19H05790, 20H00354 and 21H05233). KE and YW acknowledge support from U.S. Department of Energy, Office of Science, Basic Energy Sciences, under Early Career Award No.~DE-SC0022874. The mesoscopic \textit{ab initio} simulation used resources of the National Energy Research Scientific Computing Center (NERSC), a U.S. Department of Energy Office of Science User Facility located at Lawrence Berkeley National Laboratory, operated under Contract No.~DE-AC02-05CH11231 using NERSC award BES-ERCAP0020159.

\textbf{Author contributions:} ZRZ, PJH, and QM conceived the project. ZRZ fabricated devices, performed transport measurements, and contributed to the experimental and theoretical analysis with the help of XW, SB, and ZH under the supervision of QM and PJH. ZYZ and SC performed the theoretical modeling and band structure calculations under the supervision of EK. SYX, KE, YW, TD, NP, YZ, and LF contributed to the theoretical analysis. AG and DB performed second-harmonic generation measurements under the supervision of SYX. KW and TT grew the bulk BN single crystals. All authors discussed the results and wrote the manuscript.

\textbf{Competing interests:} The authors declare that they have no competing interests.

\textbf{Data availability:} Source data are provided with this paper. All other data that support the plots within this paper and other findings of this study are available from the corresponding authors upon reasonable request.

\setcounter{figure}{0}
\renewcommand{\figurename}{\textbf{Extended Data Fig.}}

\begin{figure*}
    \centering
    \includegraphics[width=3.375in]{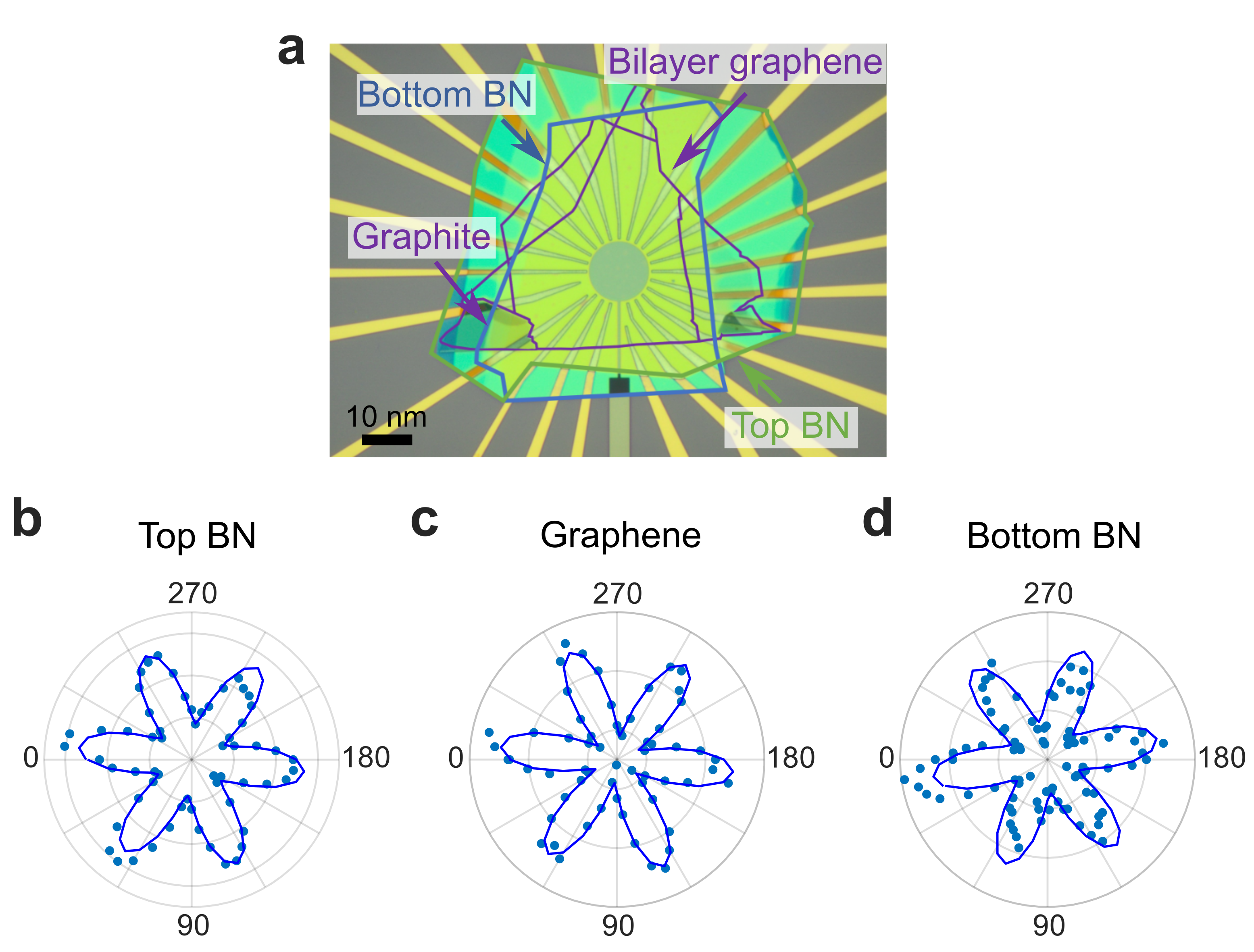}
    \caption{\textbf{Rotational alignment of Device D1.} (\textbf{a}) Optical image of the device structure. Top BN, bilayer graphene, and bottom BN are each outlined with different colors. (\textbf{b}-\textbf{d}) Second Harmonic measurement of the top BN, BLG, and bottom BN. It confirms that the BLG is closely aligned with the top BN and misaligned with the bottom BN by 15 degrees.}
    \label{SHG}
\end{figure*}

\begin{figure*}
    \centering
    \includegraphics[width=5in]{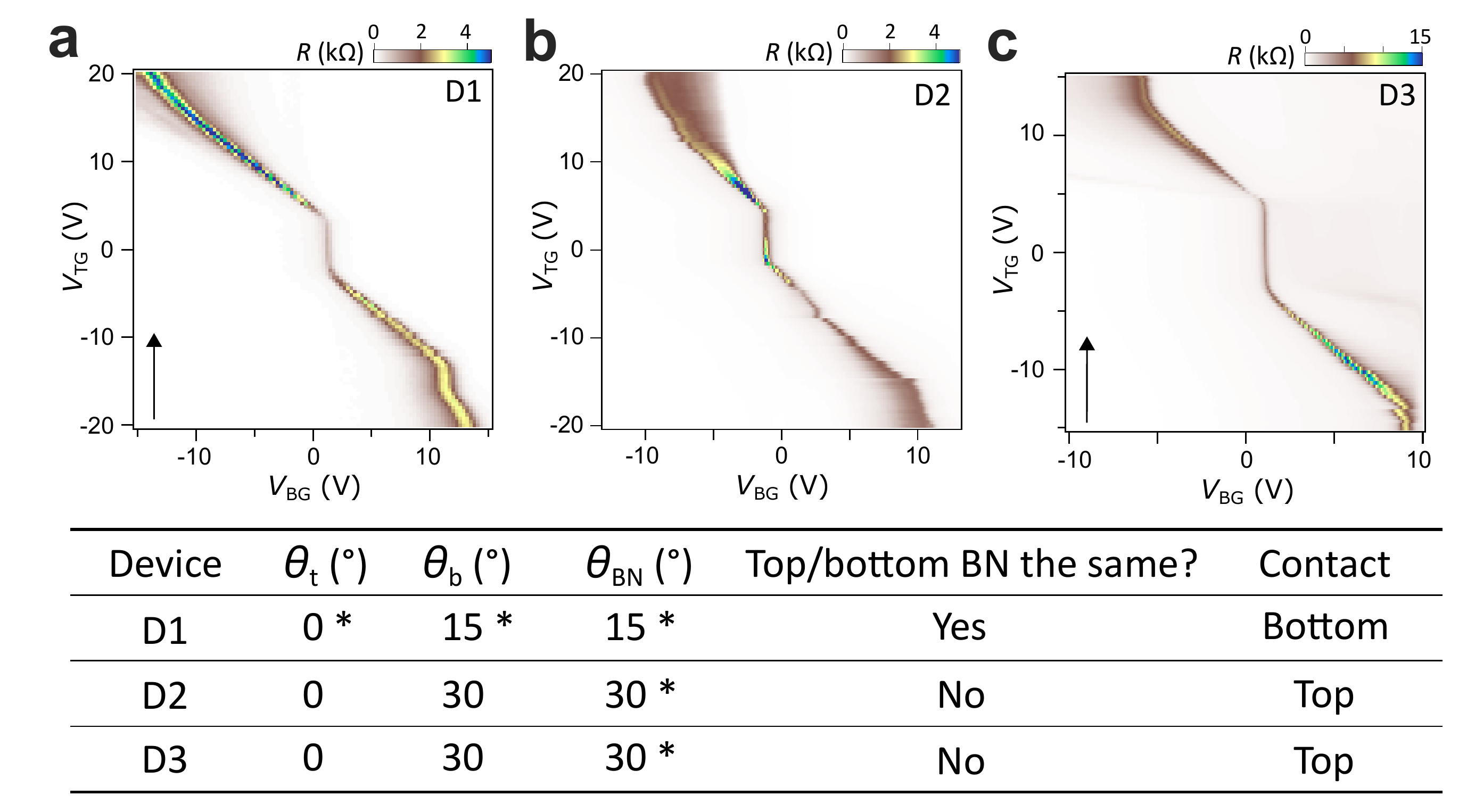}
    \caption{\textbf{Summary of hysteretic devices.} (\textbf{a}-\textbf{c}) Resistance map with the top gate in the forward direction. All devices show consistent cascade behaviors, alternating between anomalous screening and ratcheting regimes. The table summarizes the device parameters. $\mathrm{\theta_t}$, $\mathrm{\theta_b}$, and $\mathrm{\theta_{BN}}$ are relative angles of top BN/BLG, bottom BN/BLG, and top BN/bottom BN. Angles marked with $*$ sign are confirmed by SHG measurement. Angles without the $*$ sign are determined by the relative angle of the straight edges in each flake through the optical images.}
    \label{hyssummary}
\end{figure*}

\begin{figure*}
    \includegraphics[width=4in]{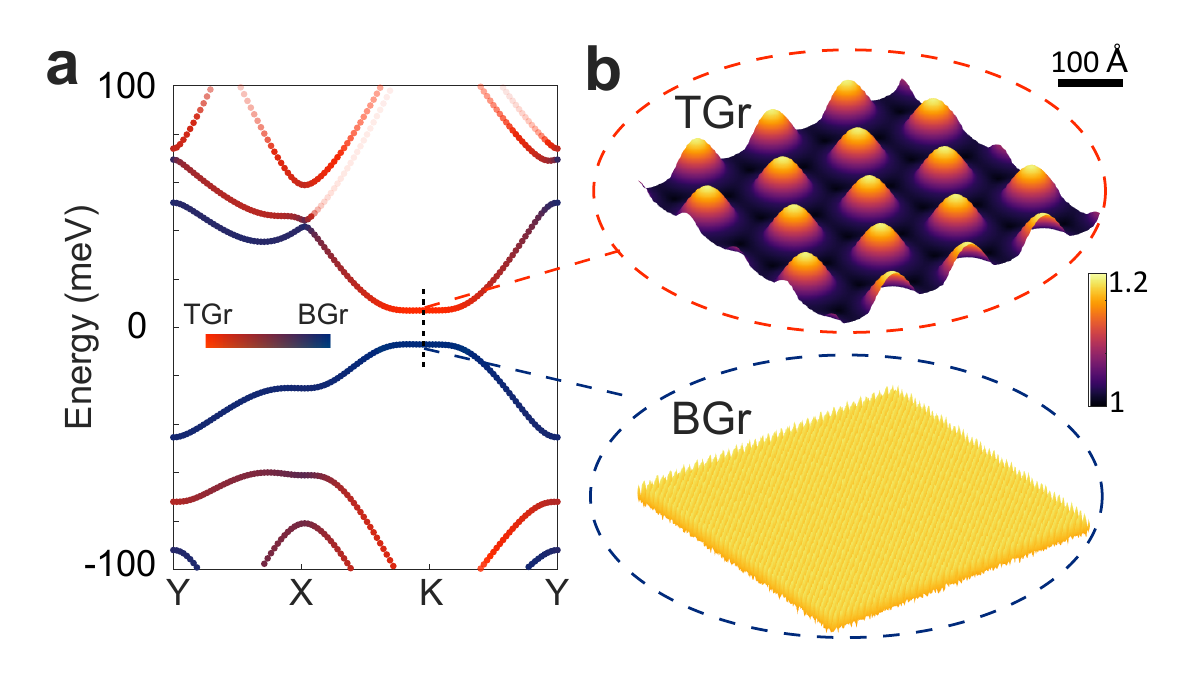}
    \caption{\figtitle{Band structure calculation and real-space wavefunction distribution}
    (\textbf{a}) Electronic band structure of a four-layer heterostructure that consists of top BN, Bernal-stacked bilayer graphene, and bottom BN. The top BN is zero-degree aligned with the BLG, while the bottom BN is misaligned with BLG by 15$\degree$. (\textbf{b}) The real-space wavefunction distribution at K point for conduction band and valance band edge at zero external electric fields.
    }
    \label{calculation}
\end{figure*}

\begin{figure*}
    \centering
    \includegraphics[width=5in]{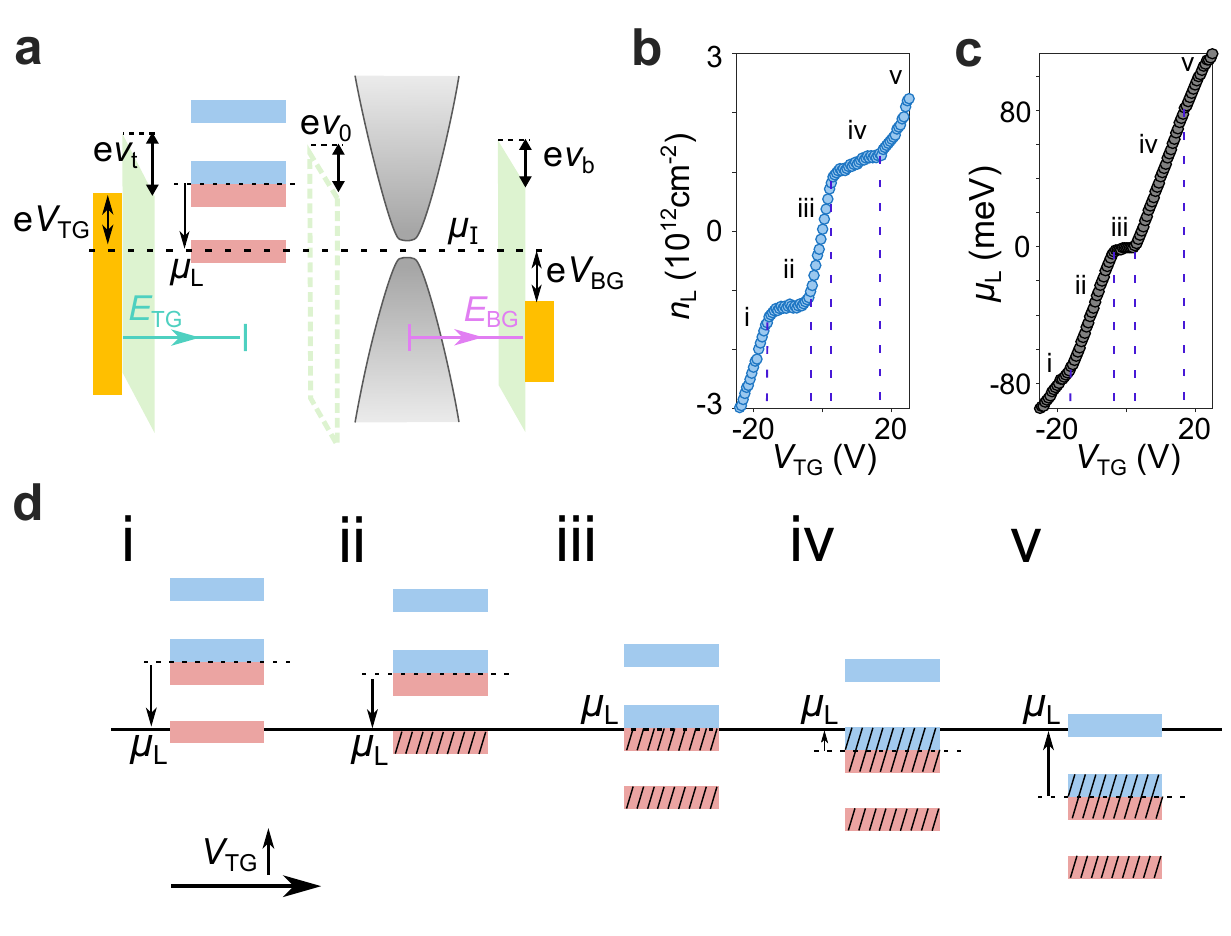}
    \caption{\textbf{Extraction of the chemical potential for the localized subsystem.} (\textbf{a}) Schematics that illustrate the relationship between the chemical potential and electrostatic potential of the localized and itinerant subsystems in response to the top gate and bottom gate voltages. (\textbf{b}) Localized charge density plotted as a function of top gate voltage. (\textbf{c}) Chemical potential of the localized charge reservoir plotted as a function of top gate voltage. (\textbf{d}) band diagram that shows the condition for regions \textbf{i} - \textbf{v} as the top gate tunes the localized electron system.}
    \label{mu}
\end{figure*}

\begin{figure*}
\centering
\includegraphics[width=3.5in]{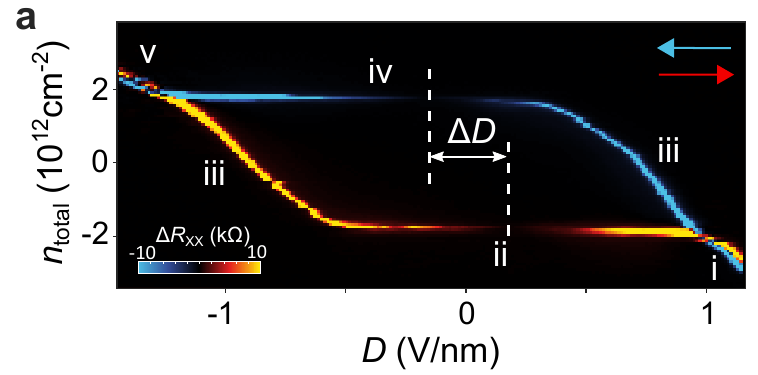}
\caption{\textbf{Hysteresis as a function of total density and displacement field.} (\textbf{a}) Four-probe resistance difference maps between the forward and backward scans along the displacement field direction. The $D$ field difference of the gapless point between the forward and backward scans is labeled as $\Delta D$.}
\label{Drange}
\end{figure*}

\begin{figure*}
    \includegraphics[width=5in]{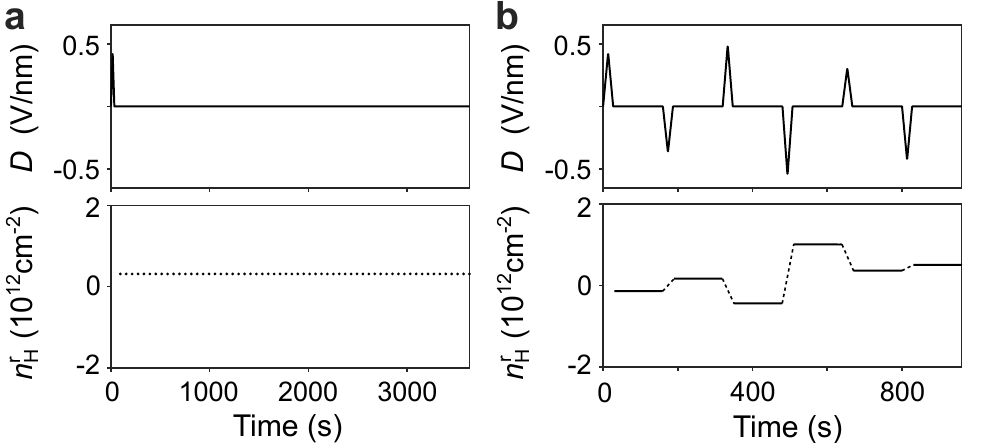}
    \caption{\figtitle{Stability and tunability of memory states.} (\textbf{a}) Remnant Hall density $n_\mathrm{H}^\mathrm{r}$ is measured as a function of time after a positive $D$ field pulse is applied. The memory state is stable over an hour and shows no sign of decay. (\textbf{b}) A series of $D$ field pulses are applied and the corresponding $n_\mathrm{H}^\mathrm{r}$ is measured. It shows that the system can be deterministically programmed to different memory states.
    }
    \label{Random}
\end{figure*}

\clearpage
\begin{figure*}
    \includegraphics[width=6.5in]{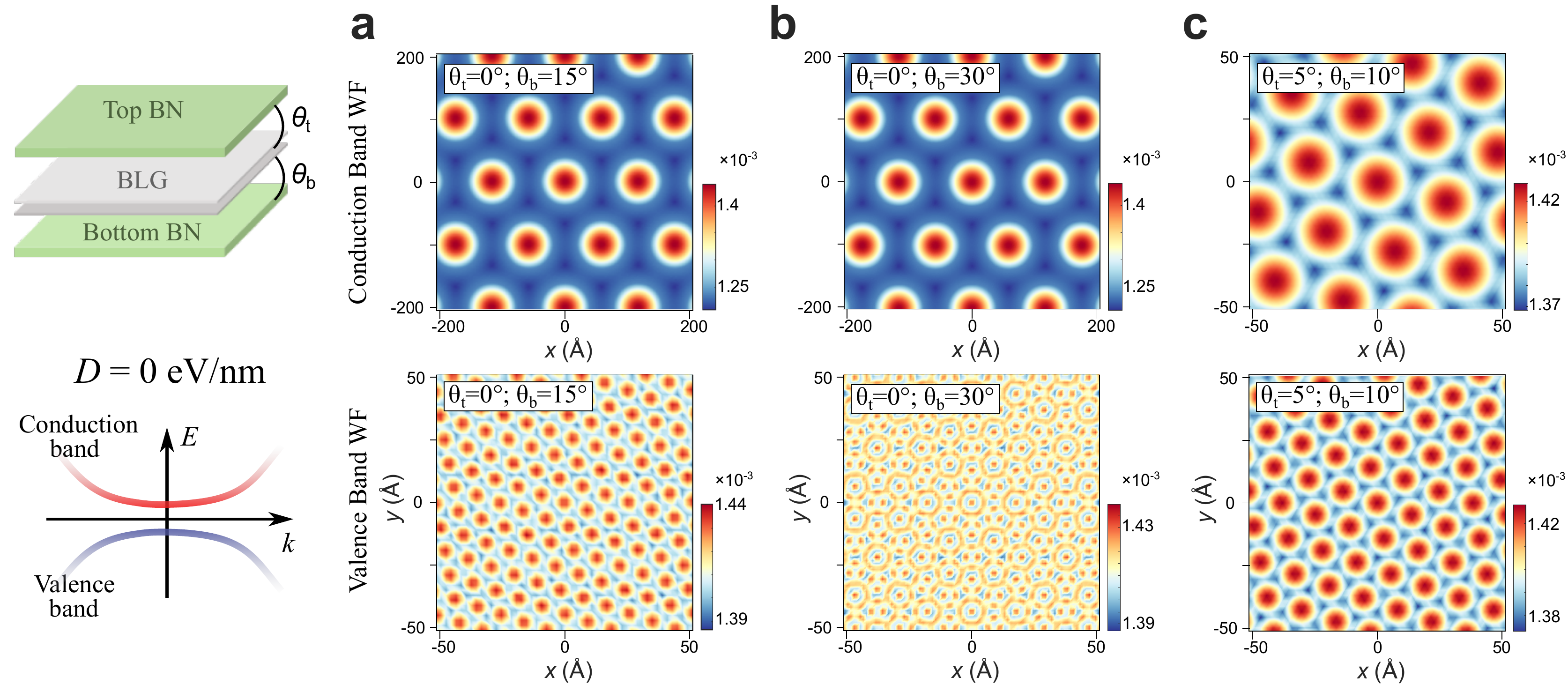}
    \caption{\figtitle{Comparison of real-space wavefunction distribution at different angle combination}
    Real-space wavefunction distribution near the conduction and valance band edge for angle combination (\textbf{a}) $\theta_{t}=0^{\circ}; \theta_{b}=15^{\circ}$ (\textbf{b}) $\theta_{t}=0^{\circ}; \theta_{b}=30^{\circ}$ (\textbf{c}) $\theta_{t}=5^{\circ}; \theta_{b}=10^{\circ}$ where $\theta_{t}$ is the angle between top BN and bilayer graphene and $\theta_{b}$ is the angle between bottom BN and bilayer graphene.
    }
    \label{WF}
\end{figure*}

\begin{figure*}
    \includegraphics[width=5in]{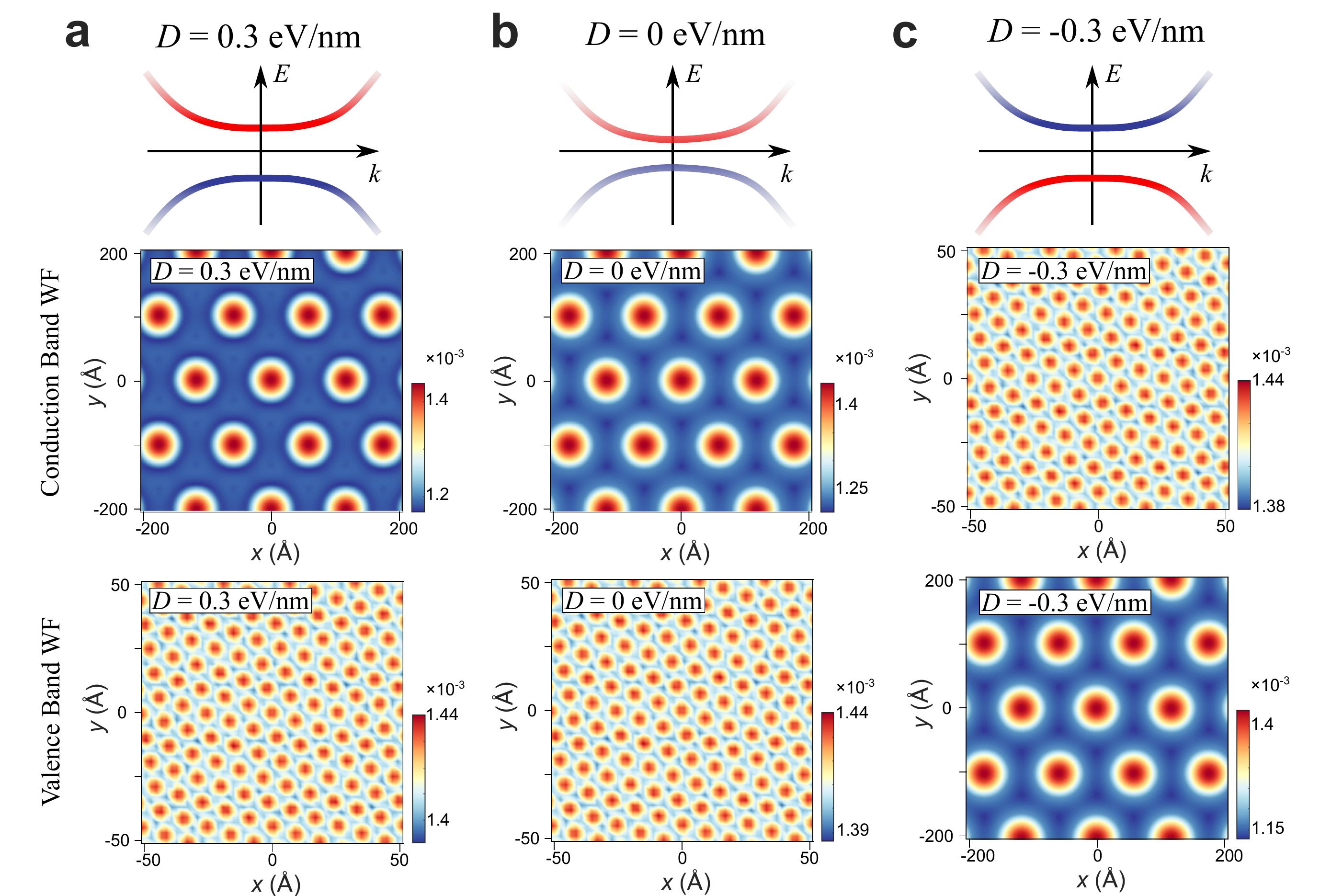}
    \caption{\figtitle{Real-space wavefunction distribution at different displacement field for $\theta_{t}=0^{\circ}; \theta_{b}=15^{\circ}$.} Real-space wavefunction distribution near the conduction and valance band edge for displacement field at (\textbf{a}) $D=0.3 $ V/nm; (\textbf{b}) $D=0$; (\textbf{c}) $D=-0.3$ V/nm.
    }
    \label{WF_Ddep}
\end{figure*}

\end{document}